\PassOptionsToPackage{table}{xcolor}
\documentclass[draft]{agujournal2019}
\usepackage{graphicx}
\usepackage{}
\usepackage{booktabs}       
\usepackage{array} 
\usepackage{cellspace} 
\usepackage{amsmath}
\usepackage{url} 
\usepackage{lineno}
\usepackage[inline]{trackchanges} 
\usepackage{soul}
\draftfalse

\journalname{arXiv}

\begin{document}

%
%


\title{A Deep Learning Earth System Model for Efficient Simulation of the Observed Climate}

%
%




\authors{Nathaniel Cresswell-Clay$^1$, Bowen Liu$^{1,2}$, Dale R.~Durran$^{1}$, Zihui Liu$^1$, Zachary I. Espinosa$^1$, Raul A. Moreno$^1$, Matthias Karlbauer$^{3}$}

\affiliation{1}{University of Washington}
\affiliation{2}{Institute of Atmospheric Physics, CAS}
\affiliation{3}{University of Tübingen}




\correspondingauthor{Nathaniel Cresswell-Clay}{nacc@uw.edu}



\begin{keypoints}
\item The coupled atmosphere-ocean Deep Learning Earth System Model (DL\textit{ESy}M) simulates 1,000 years of equilibrium climate in less than 12 hours.
\item While only trained to optimize MSE over 24 hours in the atmosphere, DL\textit{ESy}M produces realistic climatology and interannual variability. 
\item The equilibrium climate simulated by DL\textit{ESy}M has high fidelity to observed climate, competitive with state-of-the-art CMIP6 models.  
\end{keypoints}

%
%

%
%


\begin{abstract}
A key challenge for computationally intensive state-of-the-art Earth System models is to distinguish global warming signals from interannual variability. Here we introduce DL\textit{ESy}M, a parsimonious deep learning model that accurately simulates the Earth's current climate over 1000-year periods with no smoothing or drift. DL\textit{ESy}M simulations equal or exceed key metrics of seasonal and interannual variability---such as tropical cyclogenesis over the range of observed intensities, the cycle of the Indian Summer monsoon, and the climatology of mid-latitude blocking events---when compared to historical simulations from four leading models from the 6th Climate Model Intercomparison Project. DL\textit{ESy}M, trained on both historical reanalysis data and satellite observations, is an accurate, highly efficient model of the coupled Earth system, empowering long-range sub-seasonal and seasonal forecasts while using a fraction of the energy and computational time required by traditional models.
\end{abstract}

\section*{Plain Language Summary}
Machine learning (ML) based atmosphere models have recently demonstrated the ability to accurately predict weather over 10 days, and have proven superior to operational weather prediction systems based on conventional numerical methods. Many of these ML weather models, however, fail to recreate our atmosphere over longer simulations. Predicted states tend to become overly smooth, physically unrealistic, or entirely unstable. In this work we build on canonical understanding of Earth System variability by coupling an ML weather model to an ML-based model of the sea-surface. This coupling results in an Earth System model that can freely evolve without unrealistic behavior. The model is called Deep Learning Earth SYstem Model or DL\textit{ESy}M. While parsimonious in its design, DL\textit{ESy}M is capable of producing thousands of years of realistic atmosphere and ocean states including seasonal cycles, spontaneous tropical cyclones, strong winter storms, and even interannual variability. Our results are compared to those of conventional numerical Earth System models from CMIP6 demonstrating DL\textit{ESy}M’s competitive fidelity to the internal variability of the Earth System. Crucially the computational cost of running DL{\it ESy}M is significantly less compared to traditional models, making it a powerful yet affordable tool for the study of weather and climate variability.

%
%

\section{Introduction}

Accurately simulating seasonal and interannual variability in Earth System models is essential for improving sub-seasonal and seasonal forecasting (involving lead times from 2 weeks to several months) and for distinguishing global warming signals from natural variability. Recently developed atmosphere-only deep learning (DL) models provide good deterministic global weather forecasts out to about 10 days, but generally become unstable or exhibit low skill during multi-month simulations \cite{karlbauer_advancing_2024,bi_accurate_2023,lam_learning_2023}. In this work, we present a deep learning model, which couples the atmosphere with the ocean. Our model can realistically simulate the Earth's current climate over 1000-year periods with negligible drift. Our performance on key metrics of seasonal and interannual variability---such as tropical cyclone genesis and intensity, and mid-latitude blocking frequency---equals or exceeds that of historical simulations from four computationally intensive models used in the 6th Assessment Report from the Intergovernmental Panel on Climate Change (IPCC). Our model is trained on historical weather data and satellite observations to minimize forecast errors over short periods (24-h in the atmosphere and 8 days in the asynchronously coupled ocean), yet the correct seasonal and interannual variability emerges in free-running simulations. This impressive ability to learn the current climate using short-period training data sidesteps concerns about the lack of appropriate long-period datasets to train purpose-built DL models for seasonal forecasting \cite{deburgh-dayMachinelearningnumerical2023a}. Our model is an accurate, highly efficient model of the full Earth System, bridging the gap between traditional weather and climate modeling. 

Multi-year simulations with deterministic DL numerical weather prediction models have been achieved by appropriately incorporating the sea-surface temperature (SST) as lower boundary conditions. Specifying climatological SST as external forcing can improve the fidelity of some DL and DL-hybrid atmospheric models over decadal rollouts \cite{watt-meyer_ace_2023,kochkov_neural_2023} and transient SST forcing can occasion realistic atmospheric trends \cite{watt-meyer_ace2_2024}. Hybrid coupled modeling in which one component of the Earth System, represented with an DL-based model is coupled to another physics-based component model have demonstrated that the two approaches can work well together, creating stable representations of the Earth System \cite{Arcomano_etal_2023, clark_ace2-som_2024}. As an initial step with a fully DL coupled atmosphere-ocean model \cite{wang_ocean_2024} were able to approximate oceanic signatures of El Niño and equatorial waves, remaining stable for up to 6 months.

Traditional numerical models of the coupled Earth System have proved a powerful tool for scientists, facilitating the study of atmosphere-ocean interactions and feedbacks \cite{woollings_response_2012, herceg-bulic_north_2014, gupta_coupled_2007}. Their stability over long simulations and ensembling capability has also proven useful in the study of internal climate variability \cite{mann_forced_2014, marotzke_forcing_2015}. Indeed, coordinated experiments with state-of-the-art (SOTA) Earth System models through the Coupled Model Intercomparison Project (CMIP), have become one of the foundational elements of climate science \cite{Eyring_etal_2016}. These models, however, require enormous computational resources available only at large national and international centers. This large overhead constrains the use of SOTA climate models and encourages the development of novel methods that increase accessibility of high fidelity models. 

Our Deep Learning Earth SYstem Model (DL{\it ESy}M) couples a deep learning weather prediction model (DLWP) \cite{karlbauer_advancing_2024} with a  deep learning ocean model (DLOM). We examine the performance and climatology of DL{\it ESy}M over 100- and 1000-year iterative rollouts and find its fidelity to many aspects of the current climate system comparable or even superior to several leading CMIP6 models. Our simulations do not receive anthropogenic forcing and so express an equilibrium climate. Despite being an autoregressive model with no generative components, DL{\it ESy}M forecasts continue to generate sharply defined weather patterns over 730,000-step simulations. As in SOTA CMIP6 Earth System models, the coupling between the atmosphere and the ocean in DL{\it ESy}M is asynchronous, with 6-hour time resolution in the atmosphere and 2-day resolution in the ocean. In comparison to the most successful medium-range ML forecast models \cite{bi_accurate_2023,lam_learning_2023,chenFengWuPushingSkillful2023,chenFuXiCascadeMachine2023,
langAIFSECMWFDatadriven2024}, we adopt a much more parsimonious approach using an order of magnitude fewer prognostic variables at each grid point: 9 for the atmosphere and just SST for the ocean. Precipitation is not simulated directly as part of the autoregressive rollout, but rather diagnosed from the forecast fields with a separate DL module. Our training data includes fields from the ERA5 reanalysis \cite{hersbach_era5_2020}, but in contrast to previous models, we extend our set of prognostic fields to include outgoing long-wave radiation (OLR) to train on direct observations from the International Satellite Cloud Climatology Project (ISCCP) \cite{young_international_2018}. 

In this study, we will first outline important aspects of our model design, including an overview of training data. Then, we discuss the stability and fidelity of DL{\it ESy}M over millennial simulations. Afterwards, we expand our assessment of model fidelity by examining its representation of tropical cyclones and extratropical modes of variability. We also present the performance of the model in simulating the Indian Summer Monsoon (ISM). Finally we discuss El Ni\~no Southern Oscillation (ENSO) teleconnections and variability within DL{\it ESy}M simulations. 

\section{Model Design} \label{model_design}

\subsection{Architecture}

\begin{figure} 
  \centering
  \includegraphics[width=\textwidth]{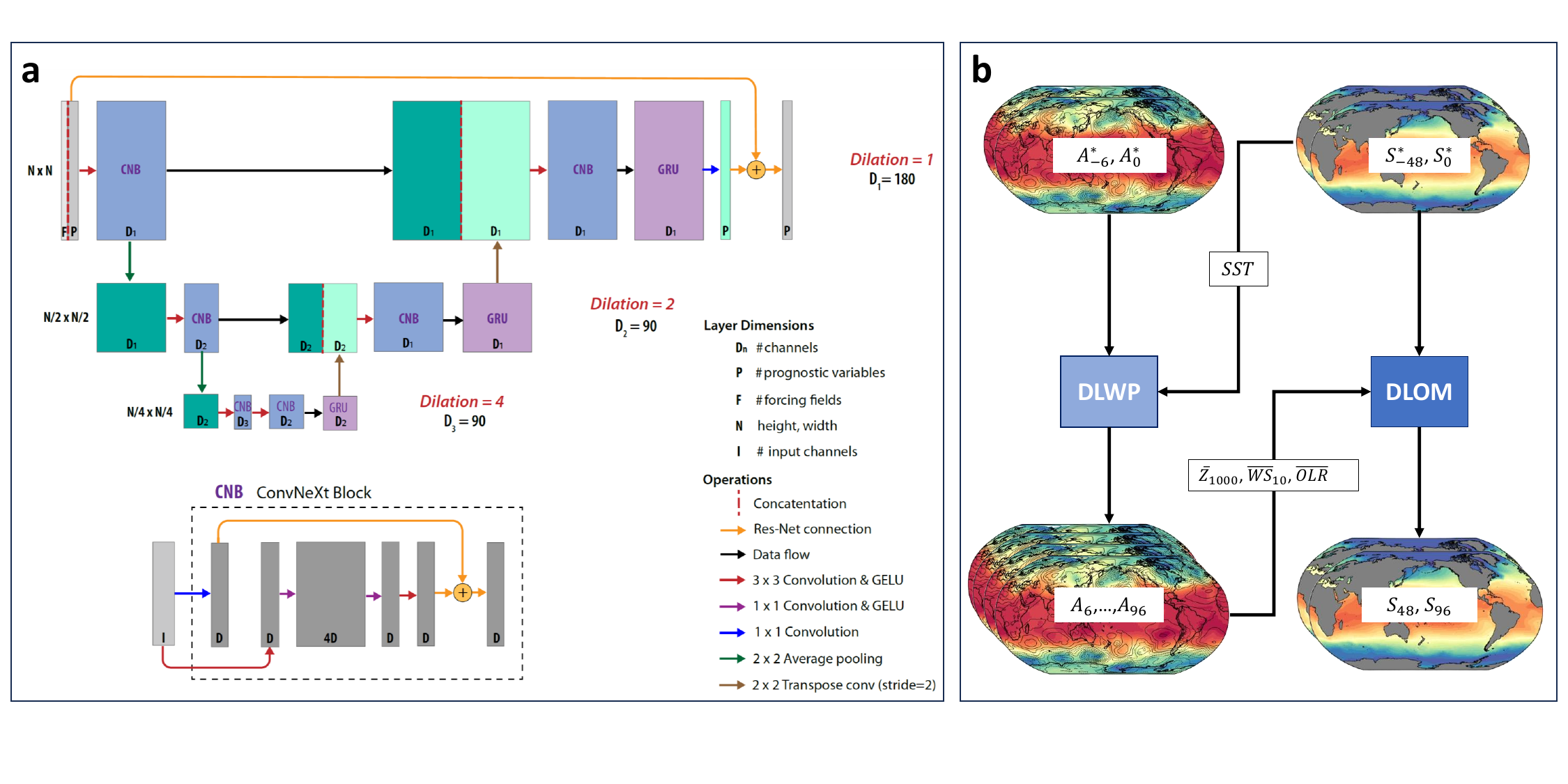}
  \caption{ {\bf Architecture and coupling of DL\textit{ESy}M}.  (a) Schematic representation of the DLWP module as a sequence of operations on layers (see legend). U-net levels are labeled by their channel depth, with $D_1=180$ and $D_2=D_3=90$ being associated with the first convolutions in each level. Each ConvNeXt block (blue) is replaced by the layers and operations shown in the inset labeled CNB, with generic embedding depths $D$ and $I$ determined by the channel depth of the input and the labeled value of $D_n$. The purple blocks labeled GRU denote convolutional Gated Recurrent Unit layers, which are implemented with $1\times 1$ spatial convolutions. Other layers evaluated by the encoder are shown as dark green, while those evaluated by the decoder are shown as light green. (b) DL{\it ESy}M coupling mechanism. Atmosphere and ocean states are denoted by $A$ and $S$, respectively, with the subscript indicating time in hours with respect to initialization at hour 0. Light blue box represents the atmospheric module (DLWP) while the darker blue represents the ocean (DLOM). Data flow and sequence of calls are given with arrows. }
  \label{arch_couple}
\end{figure}

DL{\it ESy}M uses two U-net style convolutional neural networks \cite{ronneberger2015u} in autoregessive rollouts: one simulating the atmosphere and the other, the ocean. These U-nets build on a series of successful deep learning weather prediction models  \cite{weyn_improving_2020,weyn_sub-seasonal_2021,karlbauer_advancing_2024}. 
The structure of the atmospheric module is diagrammed in Fig.~\ref{arch_couple}, with the details of the convolutional GRU \cite{ballas2015delving} omitted for simplicity. Unlike the original ConvNeXt block proposed in \cite{liu_convnet_2022}, the ConvNeXt block in \cite{karlbauer_advancing_2024} expands the latent-layer depth by a factor of four as part of a spatial convolution with a $3\times3$ kernel. Here we return to an architecture closer to that proposed in \cite{liu_convnet_2022}, where the expansion in latent-layer depth is imposed with a $1\times 1$ convolutional kernel. The resulting economy in the number of trainable weights allows us to substantially increase the latent-layer depth in each level of the U-net while reducing the total number of trainable parameters from 9.8M to 3.5M without loss of predictive skill.

The ocean module uses the same U-net architecture shown in Fig.~\ref{arch_couple}, except layer depths at each level are reduced to $D_1=90$ and $D_2=D_3=45$, and the convolutional GRU is omitted to simplify the coupling algorithm. There are roughly 0.77M trainable parameters in the ocean module. Precipitation is diagnosed with a third U-net model. This module uses the architecture in Fig.~\ref{arch_couple} with the GRU's omitted and basic convolutions in place of the ConvNeXt blocks. The final output is the accumulated precipitation between two input times, rather than an updated model state.  The precipitation module uses roughly 1.5M trainable parameters, with latent layer depths $D_1=64$, $D_2=128$, $D_3=256$. A more detailed description of the atmosphere, ocean, and precipitation modules is provided in Supporting Information. Full model configurations and weights, as well as prepared initialization data for simulations discussed in this study are published in the associated repository. 

\subsection{Atmosphere-Ocean Coupling} \label{coupling_text}

Ocean circulations evolve more slowly than those in the atmosphere so, as in traditional Earth System models, we use a longer time step for our ocean model. The DLOM updates every 4 days; each model step generates SST values at both 48-h and 96-h lead times. In contrast, the atmosphere updates every 12 hours, generating atmospheric fields at both 6-h and 12-h lead times. (See \cite{karlbauer_advancing_2024} for details about the two-in, two-out time stepping.)

To step the coupled model forward, we proceed in 96-h (4 day) increments. The mechanics of our coupling are illustrated in Fig.~\ref{arch_couple}b. To describe the mechanism of our coupled model, we use $A$ to represent the atmospheric forecast and $S$ to represent the oceanic forecast. The star operator, $\ast$, indicates observed values. Subscripts are used to represent time, expressed in hours, relative to the initialization of the simulation with hour $0$ being the time at which the forecast is initialized. 

First, we call the atmosphere model using past and initial atmospheric states, $A^{\ast}_{-6}$, $A^{\ast}_{0}$ and the initial SST, $S^{\ast}_{0}$. This call produces predicted atmospheric states $A_{6}$ and $A_{12}$. Using these predicted atmospheric states and the same $S^{\ast}_{0}$, we call the atmospheric model again. This process is repeated, using the same SST, $S^{\ast}_{0}$, throughout, until we obtain a predicted atmosphere through 96 h. The ocean states S$^{\ast}_{-48}$ and S$^{\ast}_{0}$ are then stepped forward using 10-m windspeed ($WS_{10}$), 1000-hPa height ($Z_{1000}$) and OLR as forcing from the time-averaged atmosphere states $\overline{A_{48}} = {\rm average}(A_{6},...,A_{48})$ and $\overline{A_{96}} = {\rm average}(A_{54},...,A_{96})$ to produce the predicted ocean states $S_{48}$ and $S_{96}$. This completes a full 96-h cycle, which we repeat during iterative rollouts. For a complete list of variables predicted by DL\textit{ESy}M see Supporting Information. Note that the 2-m atmospheric temperature (T$_{2m}$) is not passed to the ocean model to divorce causality from correlation, since the surface temperature over the ocean is primarily forced by the SST, not vice versa.   


\subsection{Training the Deep Learning Earth System Model}

While our ocean and atmosphere modules are coupled during inference (section \ref{coupling_text}), they are trained separately. During training, the atmosphere and ocean models learn to incorporate information about the other Earth System component by receiving reanalysis and observed values of the coupled fields. For example our atmosphere model receives ERA5 reanalysis values of SST as input in addition to prognostic variables representing the previous atmospheric state. During inference this SST input is substituted for ocean model output. In the Supporting Information we describe in detail the data, parameters, and curriculum used to train both the DLWP, DLOM, and the diagnostic precipitation module. 

\section{Weather and Climate of DL\textit{ESy}M}
\begin{figure}
\centering
\includegraphics[width=\textwidth, keepaspectratio]{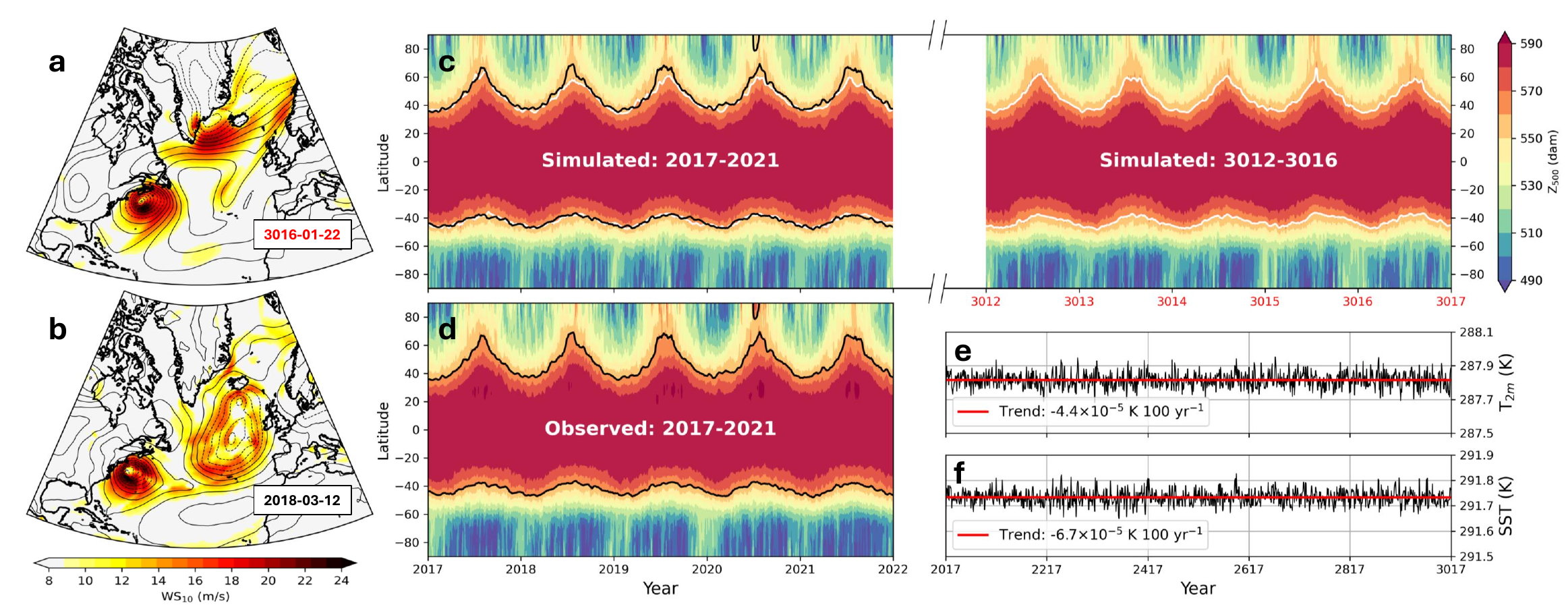}
\caption{{\bf Multi-century DL\textit{ESy}M simulations continually generate high amplitude features as well as correct representations of current global climate.} Contours of $Z_{1000}$ (black) and 10-m wind speed (color fill) for (a) a simulated ETC on January 22, 3016 and (b) observed storm on March 3, 2018.  Zonally averaged 3-day mean of $Z_{500}$ plotted as a function of time and latitude: (c) for the first and last 5 years of a recursive 1,000-year model simulation initialized on 1 January 2017 (d) corresponding $Z_{500}$ field from the ERA5 reanalysis for the years 2017-2022. Also plotted are 15-day averaged values for the 560 dam contour for the DL{\it ESy}M simulation (white line) in (c) and for ERA5 (black line) as a reference in both (c) and (d). Globally averaged annual mean (e) T$_{2m}$ and (f) SST (black) during the 1,000 year simulation with the linear fit (red) and the trend noted in each panel.}
    \label{stability}
\end{figure}

Extratropical cyclones (ETC) modulate the day-to-day cool-season weather through the passage of low and high pressure systems. A single severe ETC can cause catastrophic flooding, widespread power outages, and significant fatalities \cite{ciaran_guardian1,ciaran_guardian2}. Medium range DL weather forecast models have demonstrated skill in forecasting even severe ETCs at short lead times \cite{charlton-perezAImodelsproduce2024}, but the features forecasted by many of these models are excessively smoothed over longer lead times \cite{lam_learning_2023,langAIFSECMWFDatadriven2024}. In contrast, DL{\it ESy}M spontaneously generates realistic ETCs with sharply defined features throughout a 1000-year simulation (730,000 auto-regressive steps of the atmospheric module). In late January of simulated year 3016, for example, intense surface winds wrap around a 1000-hPa height ($Z_{1000}$) depression forming a strong ``Nor'easter" just southeast of Newfoundland Fig.~\ref{stability}a. The simulated storm demonstrates highly realistic structure when compared to an observed Nor'easter from March 12, 2018 (Fig.~\ref{stability}b).

The model's accurate reproduction of the annual cycle and negligible drift is  illustrated in Fig.~\ref{stability}c,d, which shows the first and last 5 years of a 1000-year simulation initialized on January 1, 2017, along with the ERA5 verification for the period 2017–2021. The field contoured as a function of time and latitude in Fig.~\ref{stability}c,d is zonally averaged 500-hPa geopotential height ($Z_{500}$), which is high in the tropics and lowest over the wintertime pole. There is no predictability for the precise atmospheric state beyond roughly two weeks \cite{smagorinsky1963general, mintz1964very, leith1965methods}, but the climatological annual cycle of $Z_{500}$ within the first 5 years of the 1,000-year simulation closely resemble that of the reanalysis. Impressively, this climatological seasonal cycle is maintained through the last five years of the 1000-year simulation. Similar DL{\it ESy}M results generating the same average climate were obtained for 12 shorter 100-year simulations initialized on the first day of the month throughout a full year.

Drift in global-mean values of the simulated fields is an important concern in climate modeling. Any drift must be sufficiently small that it does not mask actual changes from internal variability or anthropogenic forcing such as greenhouse-gas emissions. Despite the DL{\it ESy}M's atmospheric and ocean modules being trained separately, there is no startup transient or spin-up period in the globally averaged 2-m air temperature and SST, and the long-term drift is negligible (Fig.~\ref{stability}e,f).

\begin{figure}
  \centering  \includegraphics[width=0.9\textwidth]{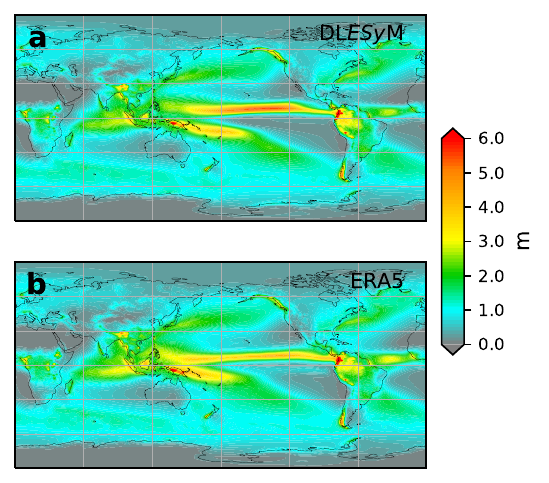}
  \caption{{\bf Climatology of global annual precipitation within DL\textit{ESy}M simulations}. Comparison of annual average precipitation (a) diagnosed from the last 41 years of our 100-year simulation, (b) from ERA5 reanalysis for 1979-2020.}
  \label{fig:annual_precip}
\end{figure}

Figure ~\ref{fig:annual_precip} shows that the annual averaged precipitation from the last 41 years of our 100-year rollout (Fig.~\ref{fig:annual_precip}a) closely matches ERA5 precipitation for the years 1979-2020 (Fig.~\ref{fig:annual_precip}b), including in regions of heavy precipitation. An exception is in the equatorial Pacific, where the DL{\it ESy}M over-estimates rainfall in the area of the Intertropical Convergence Zone. Unlike the ISCCP OLR, the ERA5 precipitation is almost exclusively model generated, and therefore subject to error. The actual precipitation is difficult to observe, although analyses such as Integrated Multi-satellitE Retrievals for GPM (IMERG) are available and could be used to train a more realistic precipitation modules. Further discussion of DL precipitation models and their development is left for further study. 

\section{Tropical Cyclones}

Tropical cyclones (TCs) have been responsible for more fatalities than any other single weather event; the 1970 Bhola TC killed 250,000--500,000 people in Bangladesh \cite{hossain1970Bholacyclone2018}. The largest TCs in the world are those in the Western North Pacific (WNP) \cite{chanGlobalclimatologytropical2015}, motivating us to compare the  climatology of TC generation over the WNP during the last 30 years of a 100-year simulation from January 1, 2017 with 30 years of ERA5 data and historical simulations from four leading CMIP6 models.  

\begin{figure}
  \centering
  \includegraphics[width=0.8\textwidth]{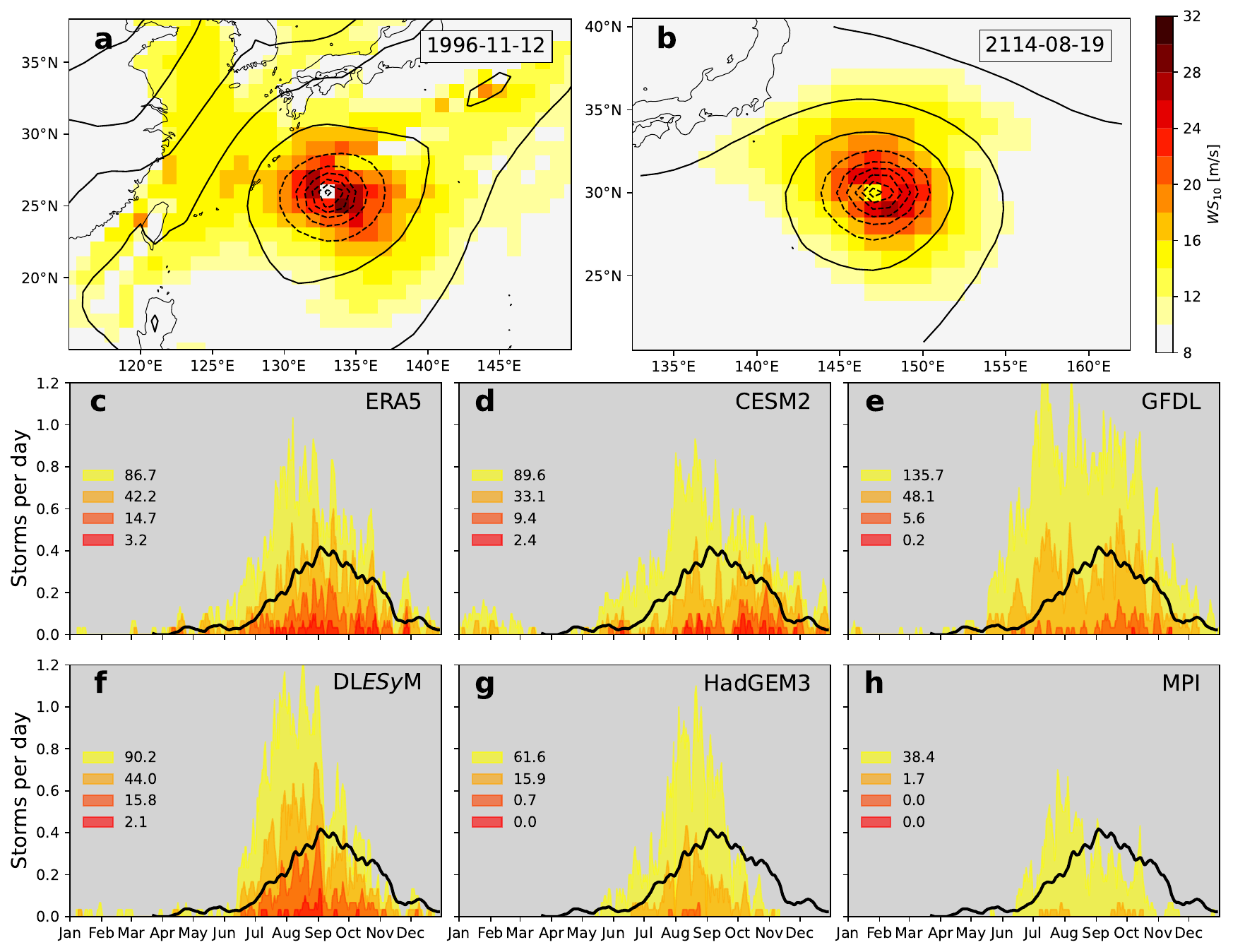}
  \caption{{\bf DL\textit{ESy}M tropical cyclone climatology in WNP compared to those of CMIP6 historical simulations}. (a) ERA5 reanalysis on November 12, 1996 and (b) DL{\it ESy}M on August 19, 2114 showing $Z_{1000}$  (black contours, c.i.=6 dkm, contours $\leq$ 0 dkm are dashed) and 10-m wind speeds (color fill). Both ERA5 and DL{\it ESy}M fields are plotted at 1$^\circ$ latitude-longitude resolution.
  (c)--(h): average frequency of TCs in storms per day from ERA5 and CMIP6 historical runs from 1985-2014, or 2085-2114 from DL{\it ESy}M. 
  Average annual number of TCs of Ranks 1-4 is noted in each panel. Black curve in (c)-(h) is the running 20-day average frequency of Rank 2 storms in ERA5.}
  \label{fig:fig2}
\end{figure}

An example of one of the top ten strongest spontaneously generated TCs near the end of the simulation (in August 2114) is compared with super typhoon Dale from the top ten TCs in the ERA5 reanalysis (on November 12, 1996) in Fig.~\ref{fig:fig2}a,b. Both TC have the same amplitude in $Z_{1000}$, although the 10-m wind speeds are weaker in the  DL{\it ESy}M simulation, perhaps because of the relatively coarse 110-km grid spacing in our model.

Similarly coarse grid spacing also imposes serious limitations in  CMIP6 models. Fig.~\ref{fig:fig2}c-h compares the TC frequency in the WNP over the 30-year period 2085-2114 in the DL{\it ESy}M simulation with the period 1985-2014 in the ERA5 reanalysis and four historical CMIP6 runs. The identification of tropical cyclones is based on the presence of a local minimum in SLP or $Z_{1000}$, an upper-level warm core, and at least 2 days of continuity along the same track, as outlined in the Supporting Information. The average number of storms per year, broken down by intensity (see Supporting Information) is also given in each panel. Except for the HadGEM3-GC31-LL, which use a 50\% coarser mesh, the CMIP6 results were obtained using roughly the same grid spacing as DL{\it ESy}M. All four CMIP6 models have difficulty capturing the correct frequency and intensity of TCs in the ERA5 reanalysis, particularly for the stronger storms. 

\begin{figure}
  \centering
  \includegraphics[width=\textwidth]{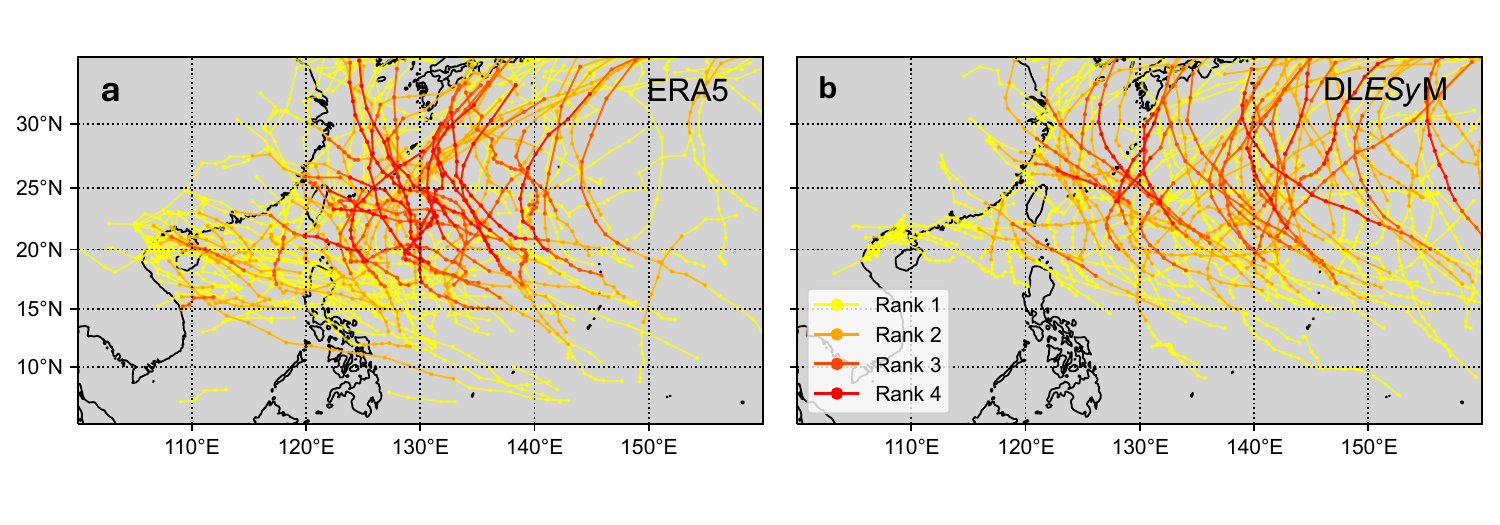}
  \caption{{\bf Tropical cyclone tracks in the Western North Pacific (WNP) are similar in the ERA5 record and DL\textit{ESy}M simulations.} Comparison of TC tracks over a 5-year period from (A) ERA5 (2010-2014), (B) from the DL{\it ESy}M rollout (2110--2114). These tracks use data at 6-h time resolution, with ranks 1--4 representing storms that, at their highest intensity exceed thresholds corresponding to approximately the 1st, 0.1st, 0.01st, and 0.001st percentile $Z_{1000}$ values in warm core systems over the western North Pacific region.  
  }
  \label{fig:TC_tracks}
\end{figure}

In contrast, the annual average number of TCs spontaneously generated by DL{\it ESy}M closely matches ERA5. The peak in DL{\it ESy}M's annual cycle does occur roughly one month early compared to the ERA5 reanalysis, a shift likely produced by a similar one-month shift in the peak SSTs simulated in this region. 
Recalling the simplicity of our DLOM, this shift might be reduced using a more complete ocean model.
Tropical cyclone tracks are similar in ERA5 and the DL{\it ESy}M simulation (Fig.~\ref{fig:TC_tracks}), although the region susceptible to TC activity extend farther to the east in our DL{\it ESy}M simulation than in observations.

\section{Atmospheric Blocking in DL\textit{ESy}M}

\begin{figure}
  \centering
  \includegraphics[width=0.8\textwidth]{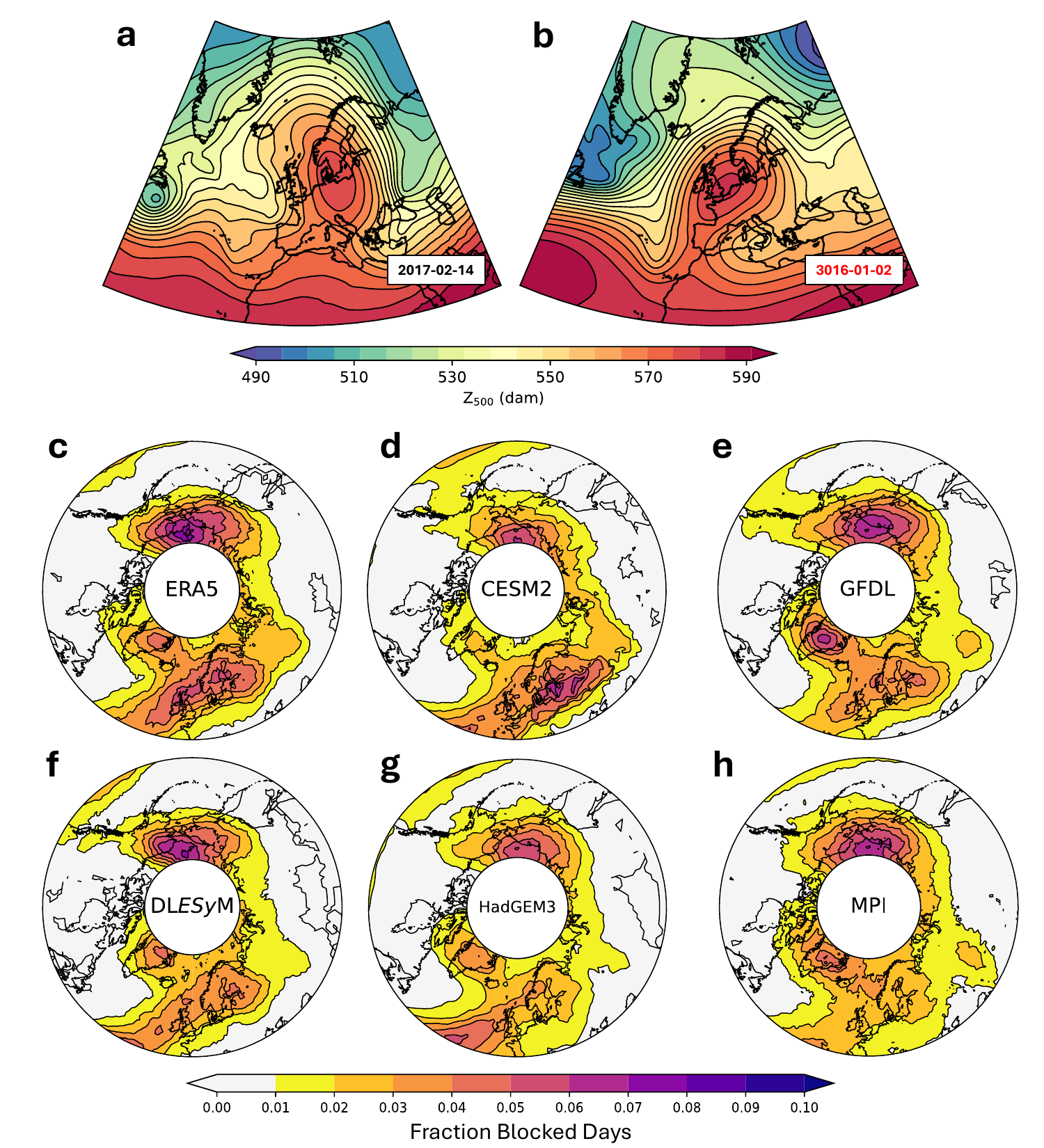}
  \caption{{\bf Representation of atmospheric blocking compared to CMIP6 historical simulations.} Contours of 500-hPa height showing blocking over Western Europe as (a) observed on February 14, 2017, (b) generated near the end of the 1000-y DL{\it ESy}M simulation on January 2, 3016. Mean frequency of northern hemisphere blocking (c) from ERA5 (1970-2010), (d, e, g, h) as in (c) computed for the same period in historical CMIP6 simulations, and (f) from the last 40 years of the 100-year DL{\it ESy}M rollout. Source models are identified in the center of each polar projection.}
  \label{fig:fig3}
\end{figure}

Blocking is produced by high-amplitude ridges that steer atmospheric flows along extended north-south trajectories leading ETCs to detour around the region underneath the ridge \cite{rex_blocking_1950,woollings_dynamical_2010,Lupo_2021}. Examples of blocking from ERA5 on February 14, 2017 and 1,000 years into the simulation on January 2, 3016 are shown by the northward extension of the $Z_{500}$ field over western Europe in Fig.~\ref{fig:fig3}a,b.
Blocking generates anomalous temperature and precipitation patterns that may include extreme cold-air outbreaks \cite{cattiaux_winter_2010,whanInfluenceAtmosphericBlocking2016}, heat waves \cite{barriopedro_hot_2011}, droughts and floods \cite{bissolliFloodingeasterncentral2011,houzeAnomalousAtmosphericEvents2011}.

Correctly capturing subseasonal and seasonal variability requires Earth System models to faithfully reproduce the frequency, spatial distribution, and amplitude of blocking events. The simulation of blocking events is a challenge for the CMIP6 models \cite{schiemann_resolution_2017,daviniCMIP3CMIP6Northern2020} and provides a good test of the climatology in the long-rollout DL{\it ESy}M simulations. Here we compare the 40-year periods 2070-2110 from a 100-year DL{\it ESy}M simulation starting on January 1, 2017 with ERA5 reanalysis and CMIP6 historical runs for the period 1970-2010.

The presence of blocking is assessed using the absolute geopotential index introduced by \cite{tibaldi_operational_1990} with parameters as defined in \cite{schiemann_northern_2020} and code adapted from the repository published with \cite{brunner_global_2017}. The time-mean blocking frequency is evaluated pointwise throughout the domain and plotted in units of blocks per day. The CMIP6 models used for these historical runs are the CESM2, GFDL-CM4, HadGEM3-GC31-LL and MPI-ESM1-2-HR.

\begin{figure}
\centering
\includegraphics[width=1.0\textwidth]{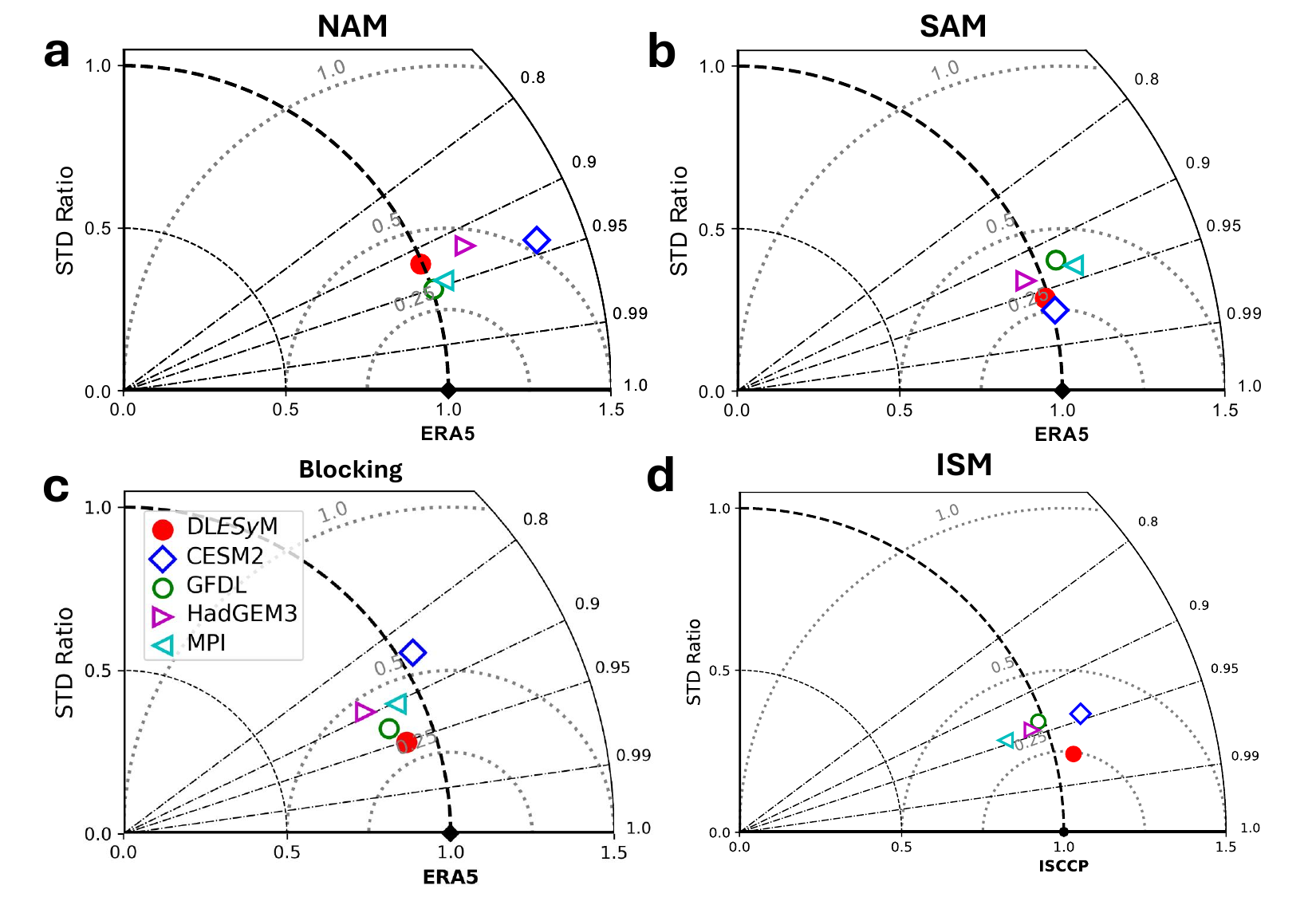}
\caption{{\bf Quantitative comparison of extra-tropical variability and the seasonal cycle of the Indian Summer monsoon between DL\textit{ESy}M and CMIP6 models.} Taylor diagrams comparing the (a) NAM (Fig.~\ref{cmip_nam_sam}a-f) (b) SAM (Fig.~\ref{cmip_nam_sam}g-l), and (c) blocking frequency (Fig.~\ref{fig:fig3}) in DL{\it ESy}M for the last 40 years of the 100-year January 1, 2017 simulation and CMIP6 historical simulations for the years 1970-2010 against the same 1970-2010 period in the ERA5 reanalysis.  The spatio-temporal OLR signatures plotted in Fig.~\ref{fig:fig4} of the Indian summer monsoon from ISCCP are compared in (d) for years 2085-2114 of the DL{\it ESy}M simulation with CMIP6 historical runs for 2085-2114. Points are plotted in a cylindrical coordinate system, with radial distance equal to the ratio of the standard deviation of the field under evaluation to the standard deviation of the target field.  The angle at which a point is plotted is given by the inverse cosine of the pattern correlation between the field under evaluation and the target. The centered pattern RMSE  is given by the distance between a point and the target point on the abscissa.
{\it Compared with the collection of four CMIP6 models, the RMSE for DL{\it ESy}M ranks 3rd out of 5 for the NAM, 2nd out of 5 for the SAM, and first for blocking and the ISM seasonal cycle.}}
    \label{Taylor_diag}
\end{figure}

DL{\it ESy}M closely approximates the spatial distribution of blocking frequencies computed from ERA5,  just slightly underestimating its frequency over Northern Europe, and the Chuckchi sea (Fig~\ref{fig:fig3}c,f). The DL{\it ESy}M result visually appears to match the pattern in ERA5 better than the corresponding results from the four CMIP6 models shown in Fig.~\ref{fig:fig3}d,e,g,h.  The performance of each model against ERA5 is quantified in the Taylor diagram in Fig.~\ref{Taylor_diag}c.

Taylor diagrams offer a visual assessment of several quantitative aspects of fidelity between simulated and observed fields \cite{taylorSummarizingMultipleAspects2001, rochfordskillmetrics}. Points are plotted on the Taylor diagram in a cylindrical coordinate system. The radial distance at which a point is plotted is the ratio of the standard deviation of the field under evaluation to the standard deviation of the ERA5 target field. The angle at which the point is plotted is given by the inverse cosine of the pattern correlation between the field under evaluation and the ERA5 target. The ERA5 target point is plotted where it would appear if it were the field subjected to evaluation: on the abscissa at a radial distance of unity. The centered pattern RMSE is given by the distance between the point for the field under evaluation and ERA5 target point on the abscissa. 

As shown in Fig.~\ref{Taylor_diag}c, the pattern correlation and centered RMSE for the DL{\it ESy}M blocking climatology are superior to those for all four CMIP models.  DL{\it ESy}M's ratio of the standard deviation of blocking frequency to that in ERA5 is slightly worse than in the CESM2, but similar or superior to that in the other CMIP6 models.

\section{Annular Modes}

\begin{figure}
\centering
\includegraphics[width=0.7\textwidth, keepaspectratio]{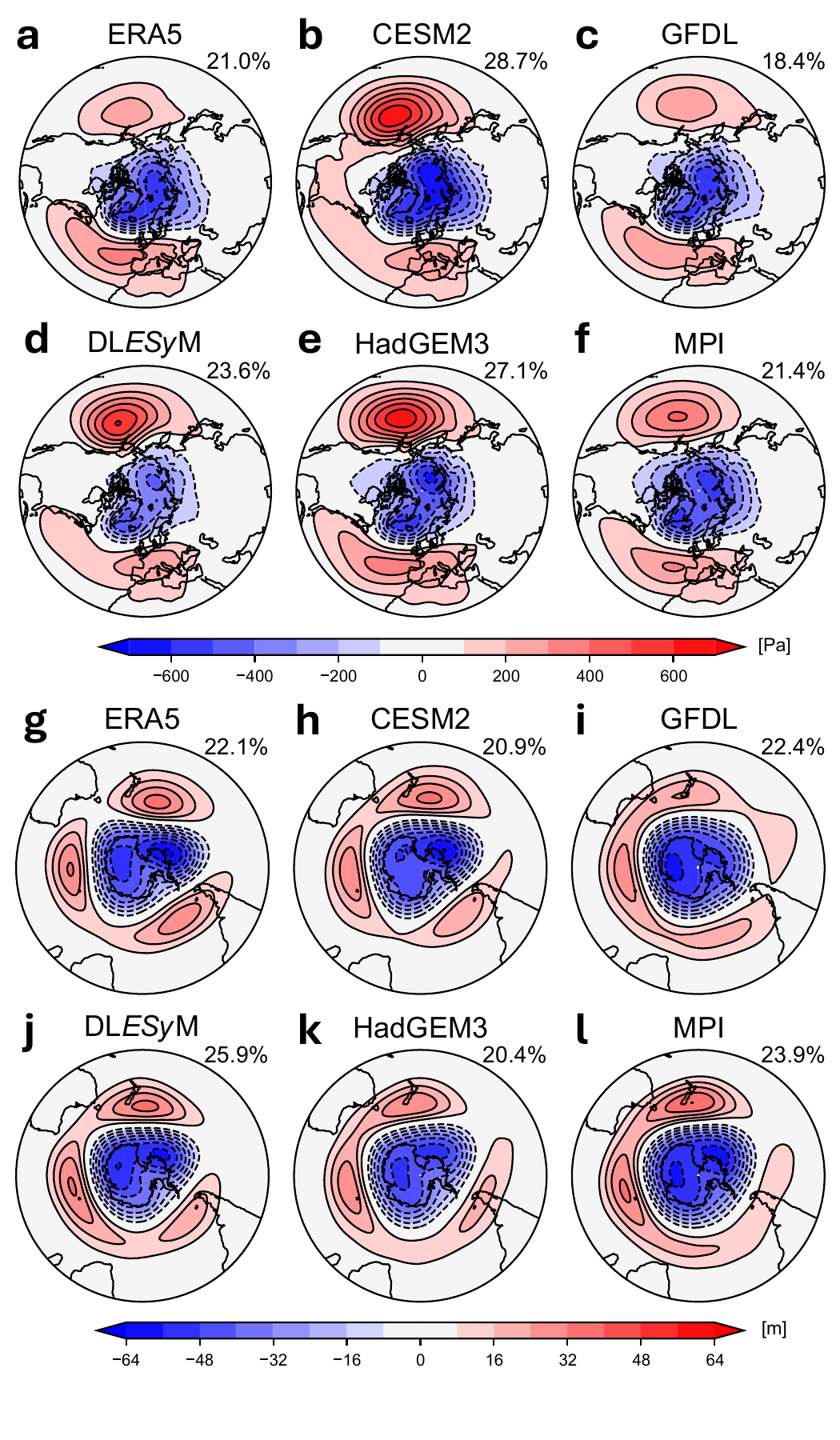}
\caption{{\bf Comparison of leading modes of extratropical variability between DL\textit{ESy}M and CMIP6 models.} Leading empirical orthogonal function (EOF) of NH variability, the Northern Annular Mode (NAM) in (a) ERA5 $Z_{1000}$ during 1970-2010; (b) SLP simulated by CESM2 for CMIP6 historical experiment; (c,e,f) as in (b) except using GFDL-CM4, HadGEM3-GC31-LL, and MPI-ESM1-2-HR respectively. (d) shows same analysis as (a) evaluated over the last 40 years of the 100 year DL{\it ESy}M rollout. Leading EOF of SH $Z_{500}$ variability, the Southern Annular Mode (SAM), in (g) ERA5 during 1970-2010; (h) simulated by CESM2 for CMIP6 historical experiment; (i, k, l) as in (b) except using GFDL-CM4, HadGEM3-GC31-LL, and MPI-ESM1-2-HR respectively. (j) shows same analysis as (g) evaluated over the last 40 years of the 100 year DL{\it ESy}M rollout. Percentage of explained variance its shown in the upper right of each panel.}
    \label{cmip_nam_sam}
\end{figure}

On seasonal and interannual time scales, the most important patterns of extratropical hemispheric-scale variability are the Northern and Southern Annular Modes (the NAM and SAM).  High magnitudes of the NAM index  are associated with statistically significant changes in the probability of impactful weather, including cold-air outbreaks and intense mid-latitude storms \cite{thompson_regional_2001}. 

The Northern Annular Mode (NAM), also commonly referred to as the Arctic Oscillation, is defined as the first Empirical Orthogonal Function (EOF) of the wintertime (November-April) monthly mean  $Z_{1000}$ anomalies north of 20\textdegree N \cite{thompson_arctic_1998, baldwin_stratospheric_2001}. The anomalies are first calculated by subtracting the climatology (seasonal cycle) at each latitude, longitude, and day of year, and then weighted by the square root of the cosine of latitude before determining the EOF. Daily values of the NAM are obtained by projecting daily anomalies onto the leading EOF patterns.

The positive phase of the spatial pattern of the NAM in ERA5 features a tri-pole pattern: with low heights in the arctic flanked by high heights over the North Pacific and the Atlantic spreading into Western Europe (Fig.~\ref{cmip_nam_sam}a). The corresponding anomaly centers are similarly arranged in the DL{\it ESy}M climatology, although  the heights are too strong over the North Pacific. Similar high biases over the North Pacific are  also present in the CESM2, MPI, and HadGEM3 climatologies. The percentages in the upper right corner in all panels in Fig.~\ref{cmip_nam_sam} show the percent variance within the full field described by each mode. DL{\it ESy}M, GFDL, and MPI show contributions from their NAM similar to that of ERA5, while CESM2 and HadGEM3 overestimate the contribution. 

The Southern Annular Mode (SAM) is calculated similarly to the NAM but for year-round monthly mean $Z_{500}$ anomalies south of 20\textdegree S. $Z_{500}$ is used instead of $Z_{1000}$ to avoid reduction-to-sea-level errors over the high terrain in Antarctica. Maps of this analysis are shown in Fig. ~\ref{cmip_nam_sam} g--l. DL{\it ESy}M expresses a realistic SAM with similar spatial structure and magnitude to the leading mode within ERA5. There is slightly less amplitude in the DL{\it ESy}M-simulated high centered over the Falkland Islands than the ERA5 record. Similar underestimation is visible in the CMIP6 models as well.

The overall performance of the DL{\it ESy}M and CMIP6 models are compared against the ERA5 NAM in the Taylor diagram in Fig.~\ref{Taylor_diag}a.  Both the GFDL and MPI models are superior to DL{\it ESy}M, which in turn scores better than the other two CMIP6 models.  The Taylor diagram for the SAM (Fig.~\ref{Taylor_diag}b) shows DL{\it ESy}M clearly outperforms three of the four CMIP6 models, only lagging behind CESM2.

\section{Indian Summer Monsoon}

\begin{figure} 
  \centering
  \includegraphics[width=0.8\textwidth]{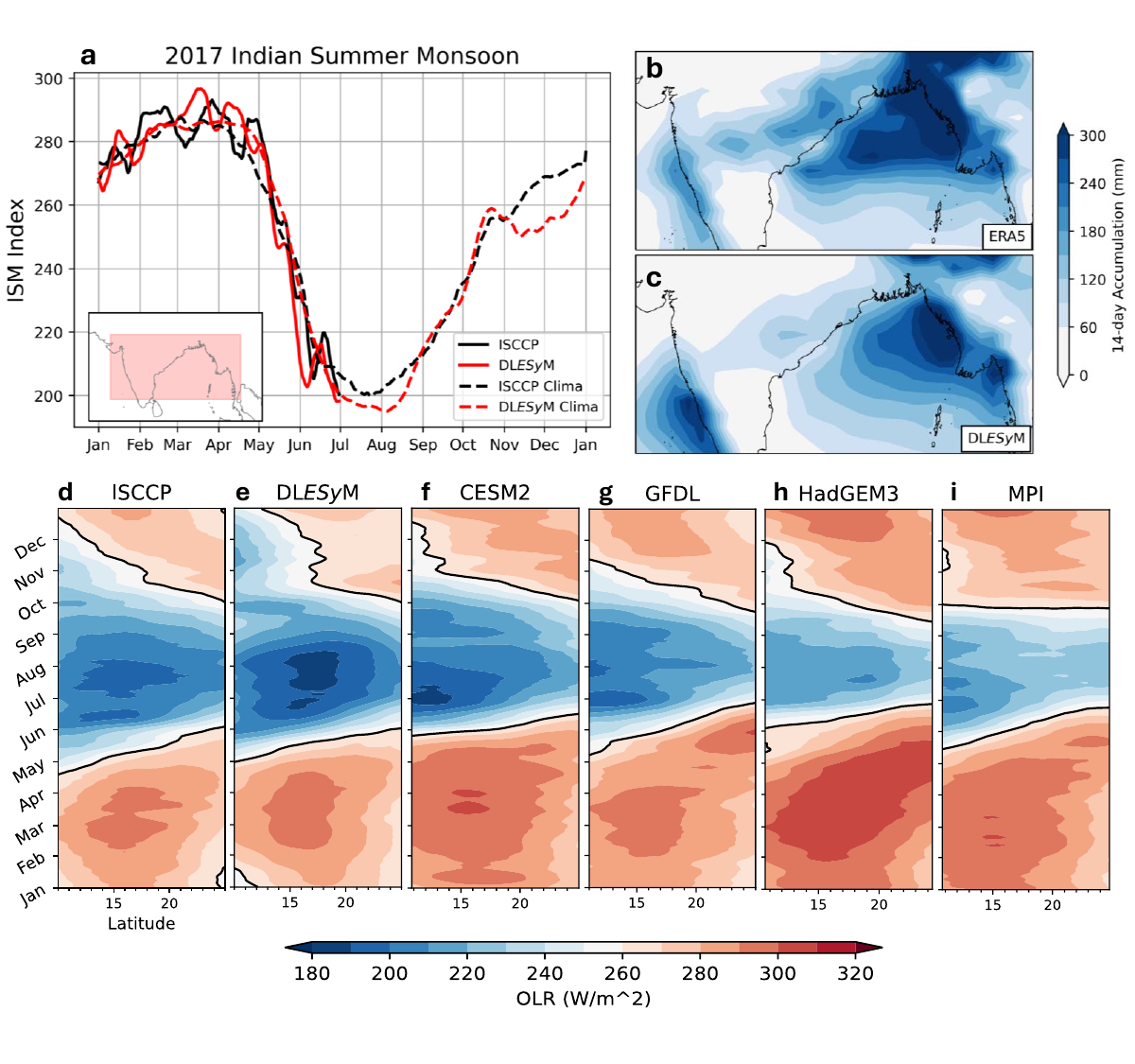}
  \caption{{\bf DL{\it ESy}M's simulation of satellite-observed cloud-modulated OLR is superior to CMIP6 models over the Indian Summer Monsoon cycle.} ISM forecast and climatology in DL{\it ESy}M. (a) Time series of forecasted 2017 ISM index (solid lines) and 30-year climatological index (dashed lines) for DL{\it ESy}M (red) and ISCCP (black). ISM index is the  OLR averaged over the red box on inserted map. 14-day accumulated precipitation beginning June 12, 2017 for (b) ERA5 and (c) DL{\it ESy}M 6-month forecast initialized January 1, 2017. Annual cycle of north-south progression of OLR  averaged zonally across the red box for 1985-2014 from (d) ISCCP observations, (f)--(i) CMIP6 historical simulations and (e) years 2085-2114 of the DL{\it ESy}M simulation.}
  \label{fig:fig4}
\end{figure}

The Indian summer monsoon (ISM) is a dominant feature in the annual cycle of tropical weather and has a crucial impact on agriculture. Increases or decreases of at least 10\% in ISM rainfall have been associated with average deviations of $\pm 10$\% in foodgrain production \cite{parthasarathyRegressionmodelestimation1988}. First, we characterize the ISM by its signature in OLR, which in contrast to precipitation, can be easily observed by satellite.  The 30-year mean annual cycle of the ISM index (after \citeNP{wangChoiceSouthAsian1999}; details provided in Supporting Information). Over the period 1985-2014 the ISM in the ISCCP data closely matches that for years 2085-2114 in the 100-year DL{\it ESy}M simulation through the start of November, after which the DL{\it ESy}M's ISM index increases more slowly than in ISCCP (Fig.~\ref{fig:fig4}a). 

The northward march of zonally averaged ISM convective cloud systems (low OLR) during onset, and their subsequent retreat to the south is plotted for 1985-2014 ISCCP data and 2085-2114 DL{\it ESy}M simulation in  Fig.~\ref{fig:fig4}d,e;  both show very similar patterns except in November and December, south of 17$^\circ$ where DL{\it ESy}M's OLR values are too low over southern India and the adjacent Indian Ocean. Similarly, the four CMIP6 models have difficulty precisely capturing the annual cycle of the ISM in their simulated OLR fields (Fig.~\ref{fig:fig4}f,g,h,i). A quantitative comparison of each model's representation of the ISM is shown as a Taylor diagram (Fig.~\ref{Taylor_diag}d). DL{\it ESy}M's ISM climatology of meridionally resolved monsoon onset has superior pattern correlation and centered RMSE to that of the four CMIP6 models when compared to ISCCP observations.

Turning from climatologies to a seasonal forecast, the OLR fields for a six-month DL{\it ESy}M forecast beginning January 1, 2017 are compared against observations up to the end of the available ISCCP data record. The forecast for the ISM index closely follows ISCCP observations, although much of this skill is derived from accurate representation of the strong climatological seasonality (Fig.~\ref{fig:fig4}a). Two-week averaged regional precipitation predicted at the end of the 6-month DL{\it ESy}M forecast is similar to the ERA5 verification (Fig.~\ref{fig:fig4}b,c). Further comparison of the DL{\it ESy}M forecast and ERA5 precipitation during the 2017 monsoon onset is included in Supporting Information.

\section{El Ni\~no  \label{sec:enso}}

\begin{figure} 
  \centering
  \includegraphics[width=\textwidth]{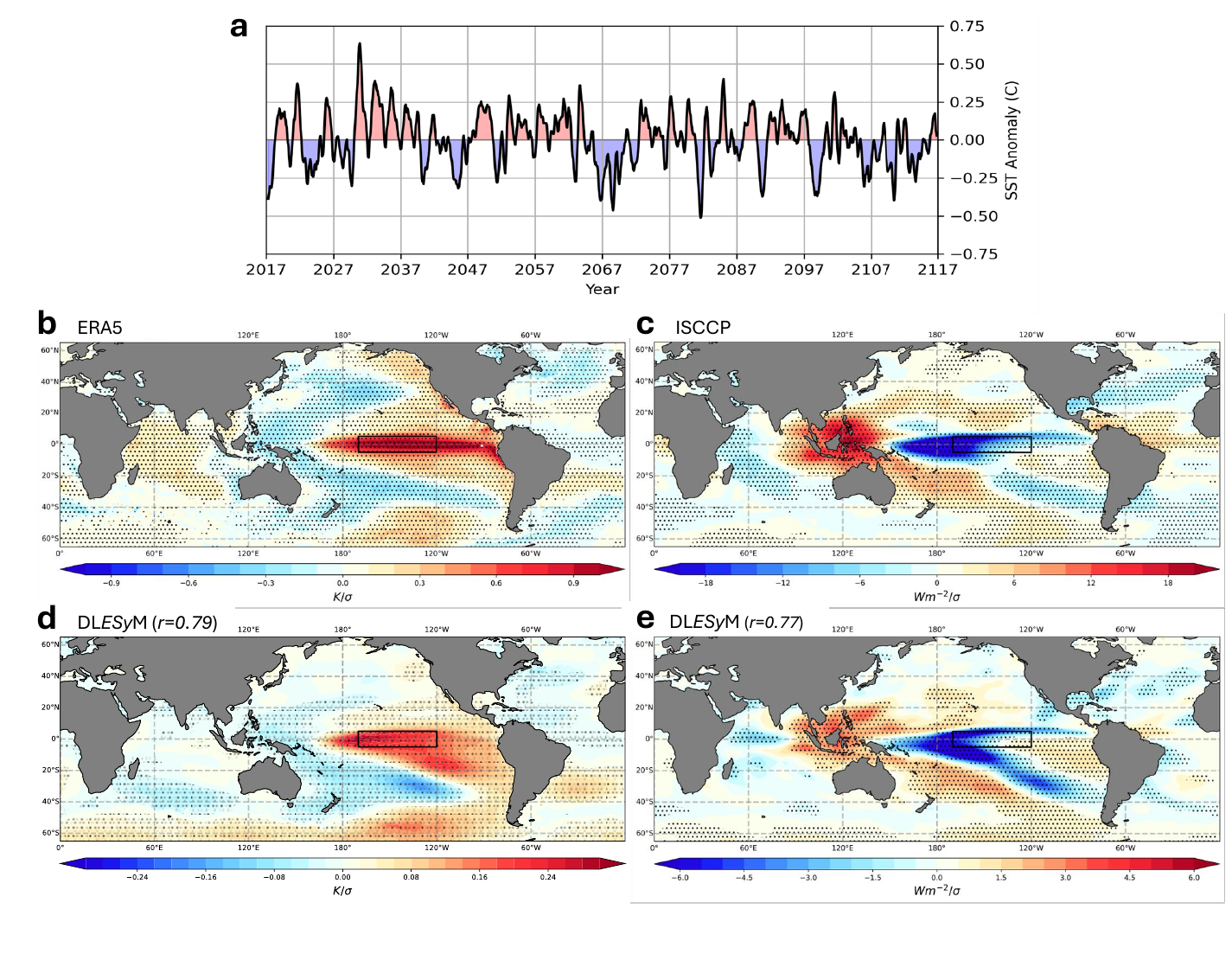}
  \caption{ \textbf{El Ni\~no Southern Oscillation variability and teleconnections in DL\textit{ESy}M.} (a) Ni\~no 3.4 index (5N-5S, 170W-120W; black box) computed from the perturbation of the equatorial SSTs from the monthly climatology in the 100-year rollout.  
  Anomalies of (b) ERA5 SST, (c) ISCCP OLR, (d) DL{\it ESy}M SST, and (e) DL{\it ESy}M OLR  regressed onto their respective standardized Ni\~no 3.4 time series. Stippling indicates statistical significance (p $\leq$ 0.05) in similarity to the verification, computed from a 2-tailed p-value associated with Pearson correlation coefficient. Correlations between modeled and observed regression maps are given in parenthetical $r$ values beside subplot titles d and e.}
\label{fig:EN1}
\end{figure}

The El Niño Southern Oscillation (ENSO) is the leading mode of interannual variability and impacts global weather patterns including precipitation over North America \cite{ropelewski1986north}, the duration of the South Asian Monsoon \cite{goswami2005enso}, and the frequency of Atlantic hurricanes \cite{grayAtlanticSeasonalHurricane1984}. 
DL{\it ESy}M spontaneously generates temporal variations in the Niño 3.4 region (5N–5S, 170W–120W) comparable to observations, albeit weaker in magnitude (Fig.~\ref{fig:EN1}a). There is no tendency for the model to drift toward a warm or cold state in the equatorial Pacific.  Instead, weak El Niños or La Niñas with magnitudes of the Niño 3.4 index exceeding 0.5 develop with a period of roughly 4 years, consistent with observations. 

To assess patterns of spatial variability, SST and OLR anomalies for observations (Fig.~\ref{fig:EN1}b,c) and model output (Fig.~\ref{fig:EN1}d,e) are regressed onto their respective standardized Niño 3.4  times series. Maps of these regression coefficients, interpreted as the SST and OLR anomalies associated with a positive one standard-deviation of the Niño 3.4 index, reveal that DL{\it ESy}M generates patterns of variability remarkably consistent with observations. Associated with positive Niño 3.4 anomalies, DL{\it ESy}M exhibits widespread warming and cooling in the eastern and western Pacific, respectively, and a weakening of the climatological west-east equatorial SST gradient. Coincident with this pattern of SST changes is a decrease in OLR in the central Pacific consistent with high cloud-tops, deep convection,  and a shift in the Walker circulation.  Positive Niño 3.4 anomalies are also associated with an increase in OLR in the western Pacific, characteristic of subsidence and reduced cloud cover. Moreover, the similarity between the DL{\it ESy}M and observed response pattern of SSTs and OLR in remote ocean basins (i.e. Atlantic and Southern Oceans) suggests DL{\it ESy}M may capture the remote atmospheric teleconnections associated with ENSO. 

Despite the low amplitude in DL{\it ESy}M’s ENSO, these results strongly suggest that, with modest improvements, DL{\it ESy}M can be a powerful new tool for studying ENSO variability. Recall that the ocean module in DL{\it ESy}M is quite basic, predicting only SSTs. A more complete model capable of capturing atmosphere-ocean interactions, thermal damping, and upper-ocean dynamics is being developed to better capture realistic ENSO amplitudes.

\section{Conclusion}

Efforts to improve SOTA Earth System models often focus on incorporating increasingly detailed representations of additional physical processes and extending to higher spatial resolutions. As a result, using these traditional models becomes increasingly difficult for those without access to very high performance computing. Here we adopt a simpler and novel approach, asynchronously coupling a DLWP model with just 9 prognostic variables to a DLOM simulating SST, both using a 110-km globally-uniform mesh. The resulting DL{\it ESy}M reproduces several aspects of the current climate at least as well as leading CMIP6 models at similar spatial resolution, yet as configured for this investigation, the DL{\it ESy}M can complete a 1000-year simulation on a single NVIDIA A100 GPU in about 12\,h.  In contrast, running a 1000-year simulation with the National Center for Atmospheric Research's CESM2.1.5 model at similar spatial resolution using 1280 processing elements on their  HPE Cray EX, would take approximately 90 days.

The simulations shown here demonstrate that our DL{\it ESy}M is not subject to several limitations that have been assumed to apply to the stability and accuracy of long autoregressive rollouts of deep learning Earth System and weather forecast models  \cite{lam_learning_2023,price_2023,chattopadhyayLongtermInstabilitiesDeep2023}. In particular, the DL{\it ESy}M simulation remains stable for 730,000 atmospheric-module steps without significantly smoothing the structure of individual weather systems. Empiricism clearly underlies our data-driven machine learning approach, while playing a lesser role in traditional Earth System models.  CMIP models do, nevertheless, also include many empirically tuned parameters. For example, parameters are typically adjusted to minimize any drift in pre-industrial climate simulations remaining after an initial spin up period of several centuries. In contrast, there is no initial transience when our separately trained atmosphere and ocean models begin the coupled simulation, and drifts in DL{\it ESy}M's globally averaged 2-m air temperature and SST are negligible.  

Surprisingly, the robust long-term properties of the DL{\it ESy}M emerge after training it on loss functions that only focus on very short-term performance. Those loss functions consist of RMSE averaged over one day in the atmosphere and eight days in the ocean. No physical constraints or components of the training loss explicitly push the model toward the correct long-term behavior.

Climatologies of 30- and 40-year periods from long iterative DL{\it ESy}M simulations were compared to similar length periods from the ERA5 reanalysis, ISCCP satellite observations, and historical runs from four CMIP models.  The frequency and spatial distribution of northern-hemisphere blocking is captured by DL{\it ESy}M at least as well as by the CMIP6 models.  In the tropics, DL{\it ESy}M rollouts replicate the observed daily climatology of tropical cyclones in the western North Pacific and the Indian Summer monsoon better than the CMIP6 models. The northern and southern annular modes (NAM and SAM) are captured with fidelity similar to the CMIP6 models as measured by spatial correlation, variance, and RMSE.  Because both the NAM and SAM are associated with extreme weather, long simulations of DL{\it ESy}M can help us understand the statistics of extreme weather patterns.

One limitation of DL{\it ESy}M is that it is only suitable for simulations of the current climate.  It remains, however, a powerful tool to study internal climate variability, and offers a radical increase in the accessibility of a high fidelity Earth System model. By accurately capturing variability across timescales, DL{\it ESy}M shows great promise for use in seasonal and subseasonal forecasting (S2S). Improved S2S forecasts, in particular, have the potential for immediate societal benefit. Indeed, ensemble DL{\it ESy}M forecasts are currently being developed for this purpose but are left as the topic of future work. Forecasts of future climates will require the model to incorporate forcing from greenhouse gases and anthropogenic aerosols; further investigation is warranted to determine whether this can be achieved by adding physical constraints to the model.

%
%

\section*{Open Research Section}
All data used in this study was obtained through publicly available data repositories. ERA5 data was downloaded from ECMWF's Climate Data Store \cite{hersbach_era5_2020, copernicus_climate_change_service_era5_2018}. ISCCP data is made available by NOAA's National Center for Environmental Information initiative \cite{young_international_2018, ncei_isscp_2021}. Code required for data preparation, model training/inference, and analysis is published on GitHub (https://github.com/AtmosSci-DLESM/DLESyM). CMIP6 data was downloaded using the acccmip6 open source project \cite{acccmip6}. 

\section*{Author Contributions}

DL{\it ESy}M conceptualization: DRD, NCC. Model development: NCC, MK, ZL. DL{\it ESy}M training: NCC. Precipitation module conceptualization: DRD and RM. Precipitation module training: RM. DL{\it ESy}M Execution of rollouts: NCC. Data curation: NCC, ZL, MK, BL, RM. Analysis design: DRD, NCC, BL, ZE. Analysis execution: BL, NCC, ZE, RM. Writing: DRD, NCC, BL, ZE, RM. Revisions: NCC, DRD, ZE, MK, BL, RM, ZL.  Figures: BL, NCC, RM, ZE, DRD. Project coordination: DRD.

\section*{Acknowledgments}

We thank Mike Pritchard and David Battisti for thoughtful comments on the manuscript. We thank Zilu Meng for comments on calculation of blocking frequency. We thank the Earth System Grid Federation (ESGF) for facilitating access to CMIP6 data. This research was supported by the Office of Naval Research under grants N0014-21-1-2827, N00014-22-1-2807, and N00014-24-12528. This work was supported in part by high-performance computer time and resources from the DoD High Performance Computing Modernization Program. NCC was supported by a National Defense Science and Engineering Graduate Fellowship. BL was supported by the Joint Ph.D. Training Program of the University of Chinese Academy of Sciences, and by the National Science Foundation of China (42075052). MK was supported by the Deutscher Akademischer Austauschdienst (DAAD, German Academic Exchange Service), by the International Max Planck Research School for Intelligent Systems (IMPRS-IS), and by the Deutsche Forschungsgemeinschaft (DFG, German Research Foundation) under Germany's Excellence Strategy EXC 2064 – 390727645. ZE was supported by the U.S. Department of Energy, Office of Science, Office of Advanced Scientific Computing Research, Department of Energy Computational Science Graduate Fellowship under Award Number DE-SC0023112.  We are grateful to NVIDIA and Stan Posey for the donation of A100 GPU cards. This research was additionally supported by a grant from the NVIDIA Applied Research Accelerator Program and utilized a NVIDIA DGX-100 Workstation. Finally, this work benefited substantially from the barrier-free high quality ERA5 dataset provided by the ECMWF.

This report was prepared as an account of work sponsored by an agency of the United States Government. Neither the United States Government nor any agency thereof, nor any of their employees, makes any warranty, express or implied, or assumes any legal liability or responsibility for the accuracy, completeness, or usefulness of any information, apparatus, product, or process disclosed, or represents that its use would not infringe privately owned rights. Reference herein to any specific commercial product, process, or service by trade name, trademark, manufacturer, or otherwise does not necessarily constitute or imply its endorsement, recommendation, or favoring by the United States Government or any agency thereof. The views and opinions of authors expressed herein do not necessarily state or reflect those of the United States Government or any agency thereof.

\section*{Conflict of Interest Statement}
The authors have no conflicts of interest to declare. 

%
%


%
\nolinenumbers

\clearpage
\setcounter{page}{1}
\begin{center}
\vspace*{30mm}
{\normalsize Supporting Information for}

{{\large \textbf{A Deep Learning Earth System Model for Efficient Simulation of the Observed Climate }}}

\authors{Nathaniel Cresswell-Clay$^1$, Bowen Liu$^{1,2}$, Dale R.~Durran$^{1}$, Zihui Liu$^1$, Zachary I. Espinosa$^1$, Raul A. Moreno$^1$, Matthias Karlbauer$^{3}$}

\affiliation{1}{University of Washington}
\affiliation{2}{Institute of Atmospheric Physics, CAS}
\affiliation{3}{University of Tübingen}

\end{center}

\vspace{30mm}
\noindent \textbf{Contents of this file} \\
\vspace{2mm}
\hspace{8mm} Text S1 to S5 \\
\hspace{8mm} Figures S1 to S4 \\
\hspace{8mm} Table S1 \\

\vspace{5mm}
\noindent \textbf{Additional Supporting Information (Files uploaded separately)} \\
\vspace{2mm}
\hspace{8mm} None

\vspace{25mm}
\section*{Introduction}

\noindent Supporting information for the manuscript entitled ``A Deep Learning Earth System Model for Efficient Simulation of the Observed Climate" is contained in this compiled PDF file. There are no externally uploaded materials. Within this PDF, we include several sections of text that offer important details which will be relevant to interested experts, but is not essential for understanding the conclusions of the study. This file contains figures and a table that are used to present these details.

\clearpage
\renewcommand{\thepage}{S\arabic{page}}
\renewcommand{\thesection}{S\arabic{section}}
\renewcommand{\thetable}{S\arabic{table}}
\renewcommand{\thefigure}{S\arabic{figure}}
\renewcommand{\figurename}{Figure}
\setcounter{figure}{0}
\setcounter{section}{0}
\section{Component Model Details}

\subsection{Deep Learning Weather Prediction Model}

The DLWP training dataset is made up of four types of input: the 9 prognostic fields predicted by DLWP as well as 2 constant fields, 1 prescribed field, and 1 coupled field. Each of these inputs is interpreted by the neural net as a single channel. The constant fields are topographic height and land-sea fraction; the prescribed field is $I_{TOA}$ (which is calculated as a function of time, longitude, and latitude); the coupled field is SST. DLWP is trained on 33 years of observations (1983-2016) and validated on 1 year of observations (2016-2017). 

As in \cite{weyn_improving_2020}, the atmosphere model is trained to optimize global root mean squared error (RMSE) over 24 hours from 6-hourly data generated by two successive model steps.  Choosing a 24-h period for the loss allows the model to learn the atmospheric evolution in a complete diurnal cycle over the full globe without pushing it toward a multi-day ensemble-mean forecast.  Despite removing appropriate global mean values, and normalizing all the prognostic variables by their standard deviation, the contributions to the loss among individual variables differ by orders of magnitude; OLR and TCWV generate the largest losses, while $Z_{250}$ produces the smallest.  We scale the weight for the $i^{\rm th}$ variable, $w_{i}$, to ensure the contributions to the RMSE loss function from each variable have similar magnitude by setting $w_{i}=0.001/V_{i}$, where $V_i$ is the unscaled validation loss for the $i^{\rm th}$ variable after a 10-epoch training. Improvements achieved from this weighting are not sensitive to the exact values of $w_i$.

Here we note some important hyperparameter choices; the full model configuration appears in the code repository associated with this article. Throughout training, we use the cosine annealing leaning rate scheduler as formulated by PyTorch's \verb|optim| library \cite{NEURIPS2019_9015} over our 250 epoch training cycle. We used gradient clipping with \verb|max_norm=0.25|. The DLWP model training took roughly 4.5 days on 4 NVIDIA A100 80GB graphics processing units (GPUs). 

\subsection{Deep Learning Ocean Model} \label{training_dlom}

Like DLWP, DLOM also has four basic types of input: prognostic, constant, prescribed, and coupled. SST is the single prognostic field predicted by our ocean component; DLOM receives $F_{LS}$  as a constant field; $I_{TOA}$ as a prescribed field; and $WS_{10m}$, $Z_{1000}$ and OLR from the atmosphere. The coupled fields used to force the training of the DLOM component have been averaged over times t$_{0}$-t$_{48}$ and t$_{48}$-t$_{96}$. This treatment of the incoming atmospheric fields is designed to mimic the behavior during inference (Section \ref{coupling_text}). Notably, we do not use $T_{2m}$ as a forcing field for DLOM, despite it being a prognostic field of our DLWP model. This is done to avoid  SST values matching the $T_{2m}$ field above the ocean, when in fact it is the SST that should determine $T_{2m}$ over the ocean.

The DLOM component model is trained to optimize maritime RMSE over 192 h of prediction. Maritime RMSE is defined as the RMSE spatially weighted by $1-F_{LS}$. DLOM has 48-h resolution and so we advance 192 h from two auto-regressive calls (similar to the two calls in DLWP training). 

Training DLOM models is more sensitive than training DLWP models. To achieve best results, we employed a staged learning program with 3 phases. The program proceeded for 300 epochs, with a cosine annealing restart every 100 epochs using the best checkpoint from the previous phase.  Maximum (minimum) learing rates for each phase were 1e-4 (4e-5), 5e-5 (5e-6) and 
2.5e-5 (0), respectively.

\subsection{Deep Learning Precipitation Diagnosis\label{precip_methods}}

The input to the diagnostic precipitation module is the same two-time-level tensor of prognostic, constant, and prescribed variables as DLWP, but rather than a recurrent forecast, its output $P$ is the cumulative  6-h precipitation between the two  sequential times.   The ERA5 precipitation  over the same period is the target.  Because precipitation is highly skewed, the loss function is evaluated as in \cite{Rasp_Thuerey2021} using the log-transformed field 
\begin{align*}
    \tilde{P} = \ln\left( 1 + P/\epsilon \right)
\end{align*}
where $\epsilon = 1 \times 10^{-8}$ m.  To emphasize heavy precipitation events, the cost function, $L$, for our precipitation module is the global RMSE scaled by the exponentially transformed field,
\begin{align*}
    L=\exp{(b*\tilde{P})}\times RMSE
\end{align*}
where $b=0.4$ is a tuned hyperparameter. 

\section{Data Sources} \label{data}

\begin{table} 
    \centering
    \begin{tabular}{SlSlSlSl}
        \hline
        \textup{Variable Name}        & Symbol     & Source    & Preprocessing\\
        \hline
        \rowcolor{gray!10}
        1000-hPa geopotential height & $Z_{1000}$       & ERA5    & - \\   
        500-hPa geopotential height& $Z_{500}$       & ERA5    & - \\
        \rowcolor{gray!10}
        250-hPa geopotential height & $Z_{250}$       & ERA5    & - \\
        700--300-hPa  thickness & $\tau_{300-700}$       & ERA5    & difference between 300 and 700-hPa geopotential heights\\
        \rowcolor{gray!10}
        2-m temperature      & $T_{2m}$    & ERA5     & - \\
        850-hPa temperature      & $T_{850}$    & ERA5     & - \\
        \rowcolor{gray!10}
        total column water vapor & TCWV       & ERA5    & - \\
        10-m windspeed  & $WS_{10}$       & ERA5    & magnitude of 10-m horizontal velocity vector\\
        \rowcolor{gray!10}
        sea surface temperature  & SST       & ERA5        & land imputation \\
        outgoing longwave radiation & OLR & ISCCP & brightness conversion and OLR imputation\\
        \rowcolor{gray!10}
        land-sea fraction  & $F_{ls}$       & ERA5    & - \\
        surface elevation  & $z$       & ERA5    & - \\
        \rowcolor{gray!10}
        top of atmosphere insolation  & $I_{TOA}$       & -    & - \\
        \hline
        \\
    \end{tabular}
    \caption{{\bf Initialization fields Earth System fields for DL\textit{ESy}M.} The columns show the variable name, the abbreviated symbol for the variable, the data source, extra preprocessing steps used to prepare data for training/inference}
    \label{variable_descriptions}
\end{table}

\subsection{ERA5}

ECMWF reanalysis version 5 (ERA5) \cite{hersbach_era5_2020} is the primary dataset used for training and verification of DL{\it ESy}M, providing these prognostic fields: geopotential at 1000, 700, 500, 300 and 250 hPa, temperature at 2-m and 850 hPa, total column water vapor, and SST.  We also use two ERA5 time-invariant fields for training and inference: terrain height and land-sea fraction (called surface geopotential and land-sea mask in ERA5 database respectively).  Some of these fields require additional preprocessing using other ERA5 fields as indicated in Table~\ref{variable_descriptions}.  DL{\it ESy}M discretizes all fields using the Hierarchical Equal Area isoLatitude Pixelization (HEALPix) \cite{gorski_healpix_2005,karlbauer_advancing_2024, zonca_healpy_2019}, with the 12 primary HEALPix faces divided into  $64\times64$ cells, giving a total of 49,152 cells globally. The diagonal distance between adjacent cell centers is approximately 110 km, which in turn is approximately 1\textdegree \, of latitude at the equator.   

As part of this reprocessing, SST values over the continents are imputed using zonal linear interpolation. This coast-to-coast interpolation ensures that our convolutional operations yield reasonable values throughout our domain while minimizing the sharp gradients near the land-sea boundaries. As a possible alternative to zonally interpolated imputation, we also tested the using zonal climatological averages for SSTs over land, but this tended to produce nonphysical features in the ocean state during long rollouts. Although the full global SST is updated every step, the values over land are masked and ignored in the loss function during training.   

\subsection{ISCCP}

The International Satellite Cloud Climatology Project (ISCCP) calibrates and compiles observations from a large suite of satellites into a single dataset  \cite{young_international_2018}. Here we use the "IR Brightness Temperature" from the HXG distribution, which is available from 1983 to 2017 at 3-hour resolution on a 0.1 degree latitude-longitude mesh. The ISCCP data requires several preprocessing steps before it is ready for training. There are swaths of missing data, notably prior to 1998 over the Indian Ocean, which are filled in using ERA5's model-derived total net thermal radiation (TTR) field. To complete this imputation, IR brightness temperature is converted to outgoing longwave radiation (OLR), in units of W m$^{-2}$, by the Stefan–Boltzmann law, and downsampled  from the 0.1 degree latitude-longitude mesh to a 0.25\textdegree \, mesh to match the ERA5 grid. Then TTR, in units of J m$^{-2}$,  is scaled to match the distribution of the ISCCP brightness temperatures using the relation: 

\begin{align}
    \widehat{\mbox{TTR}}=\frac{\sigma_{\mbox{\tiny OLR}}}{\sigma_{\mbox{\tiny TTR}}}(\mbox{TTR} - \mu_{\mbox{\tiny TTR}})+\mu_{\mbox{\tiny OLR}}
\end{align}

where $\widehat{\mbox{TTR}}$ is scaled TTR, $\sigma_{\rm OLR}$ is the standard deviation of the outgoing longwave radiation, $\sigma_{\rm TTR}$ is the standard deviation of the TTR, $\mu_{\rm TTR}$ is the mean TTR, and $\mu_{\rm OLR}$ is the mean outgoing longwave radiation. Once scaled, we use the TTR to impute the outgoing longwave radiation over the missing regions of missing ISCCP data. A 2D Gaussian filter is used to smooth seams in the ISCCP observations and the imputed TTR. The result is a realistic OLR record that contains no missing data. Like ERA5 sourced data, ISCCP data was mapped onto the HEALPix mesh for training and inference. 



\section{Tropical Cyclone Detection}\label{tc_freq_methods}

\begin{figure}
\centering
\includegraphics[width=0.8\textwidth, keepaspectratio]{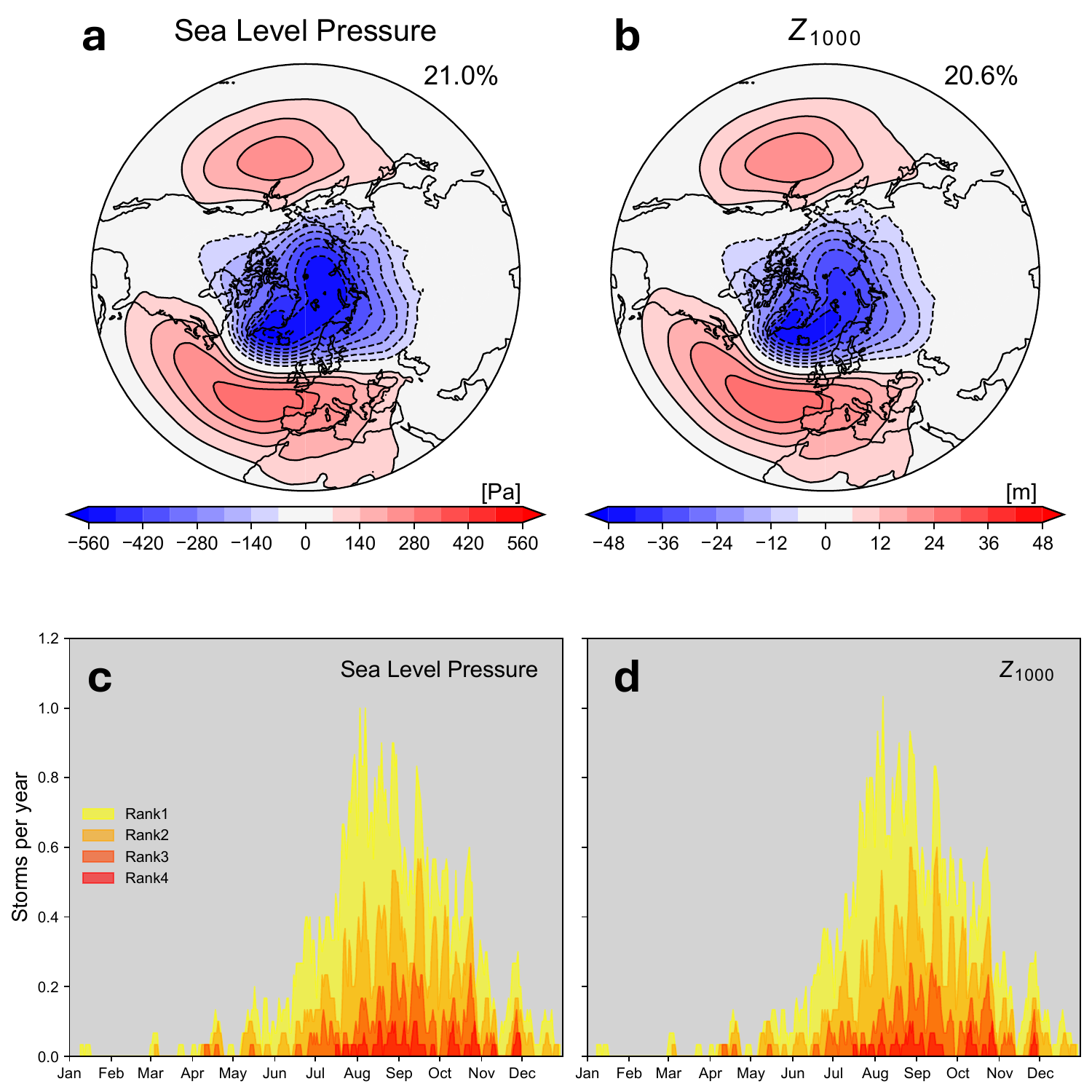}
\caption{ {\bf Correspondence between sea level pressure (SLP) and $\boldsymbol{Z_{1000}}$ in TC and NAM analyses.} Leading mode of NH variability in ERA5 evaluated using (a) SLP and (b) $Z_{1000}$. Calculated over 40 years of ERA5 record (1970-2010). Contour intervals in Pa and m, as labeled, to reveal the same pattern. TC detection algorithm applied to ERA5 using (c) SLP and (d) $Z_{1000}$. Nearly identical structure of TC frequency using the rank 1 to rank 4 thresholds noted in Supporting Information Section \ref{tc_freq_methods} justifies comparisons between DL{\it ESy}M simulations (which only predict $Z_{1000}$) and CMIP6 output (which only reports SLP).}
    \label{SLP_Z1000_comparsion}
\end{figure}

Following \cite{ullrich2021tempestextremes,zarzycki2017assessing}, this study categorizes tropical cyclones into four groups based on these three criteria:

\begin{enumerate}
    \item The local minimum in SLP drops below predefined thresholds within the western North Pacific region (100\textdegree E--160\textdegree E, 5\textdegree N--35\textdegree N). These thresholds are round numbers approximating the 1st, 0.1st, 0.01st, and 0.001st percentile SLP values. For SLP fields sampled once per day these values are 1000, 994, 985, 975 hPa. 
    
    \item The presence of a well-defined upper-level warm core. A warm core is defined as the geopotential layer thickness between 300 and 700 hPa that exceeds the seasonally varying climatological value.

    \item The tropical cyclone trajectories are continuous. The adjacent centers along the same track are separated by no more than 2 days in time and 3 degrees in spatial distance. Tracks containing fewer than 3 data points are not considered in the analysis.  Tropical cyclone tracks from ERA5 and DL{\it ESy}M are compared in Fig.~\ref{fig:TC_tracks}.
\end{enumerate}

The DL{\it ESy}M analysis employed $Z_{1000}$ to identify tropical cyclones because it does not include SLP in its state vector. Both variables are available in ERA5, which allows us to calibrate the $Z_{1000}$ values in the DL{\it ESy}M rollout for comparison with the CMIP6 historical runs. 
The difference between using SLP and $Z_{1000}$ for detecting tropical cyclones, with appropriately scaled thresholds, was negligible.  Choosing 
0, -50, -130, -230 m as the $Z_{1000}$ thresholds corresponding to those previously noted for SLP, we obtain essentially the same distribution as using SLP (Fig.~\ref{SLP_Z1000_comparsion}c,d).  The correlations between the distributions for Ranks 1 to 4 is 0.99, 0.99, 0.98, and 0.89, respectively.

\section{Annular Mode Computation}

The CMIP6 models used in this study do not have complete $Z_{1000}$ fields, and so we utilized sea-level pressure (SLP), a field not available in the DL{\it ESy}M, to calculate the NAM for the CMIP6 models.  The spatial correlation between the ERA5 NAM computed from the $Z_{1000}$ and the SLP fields is 0.99, and, when plotted with properly calibrated contour intervals, their signatures are essentially identical. Using these contour intervals for intercomparison in Fig.~\ref{cmip_nam_sam}a--f, the NAM is plotted based on the $Z_{1000}$ field for ERA5 and the DL{\it ESy}M rollout while the NAM is calculated from SLP for the CMIP6 historical runs.  To maximize compatibility, the statistics for the NAM calculated from CMIP6 data are compared against the ERA5 SLP record in the Taylor diagrams in Fig.~\ref{Taylor_diag}. 

The correlation between the NAM (SAM) index in this paper and the identical but alternatively named Arctic Oscillation (Antarctic Oscillation) index in daily time series provided by the Climate Prediction Center (CPC; https://www.cpc.ncep.noaa.gov/) for 1970–2010 (1979-2010) is 0.98 (0.93). 

\section{Indian Summer Monsoon\label{CI1_index}}

\begin{figure} 
  \centering  \includegraphics[width=0.7\textwidth]{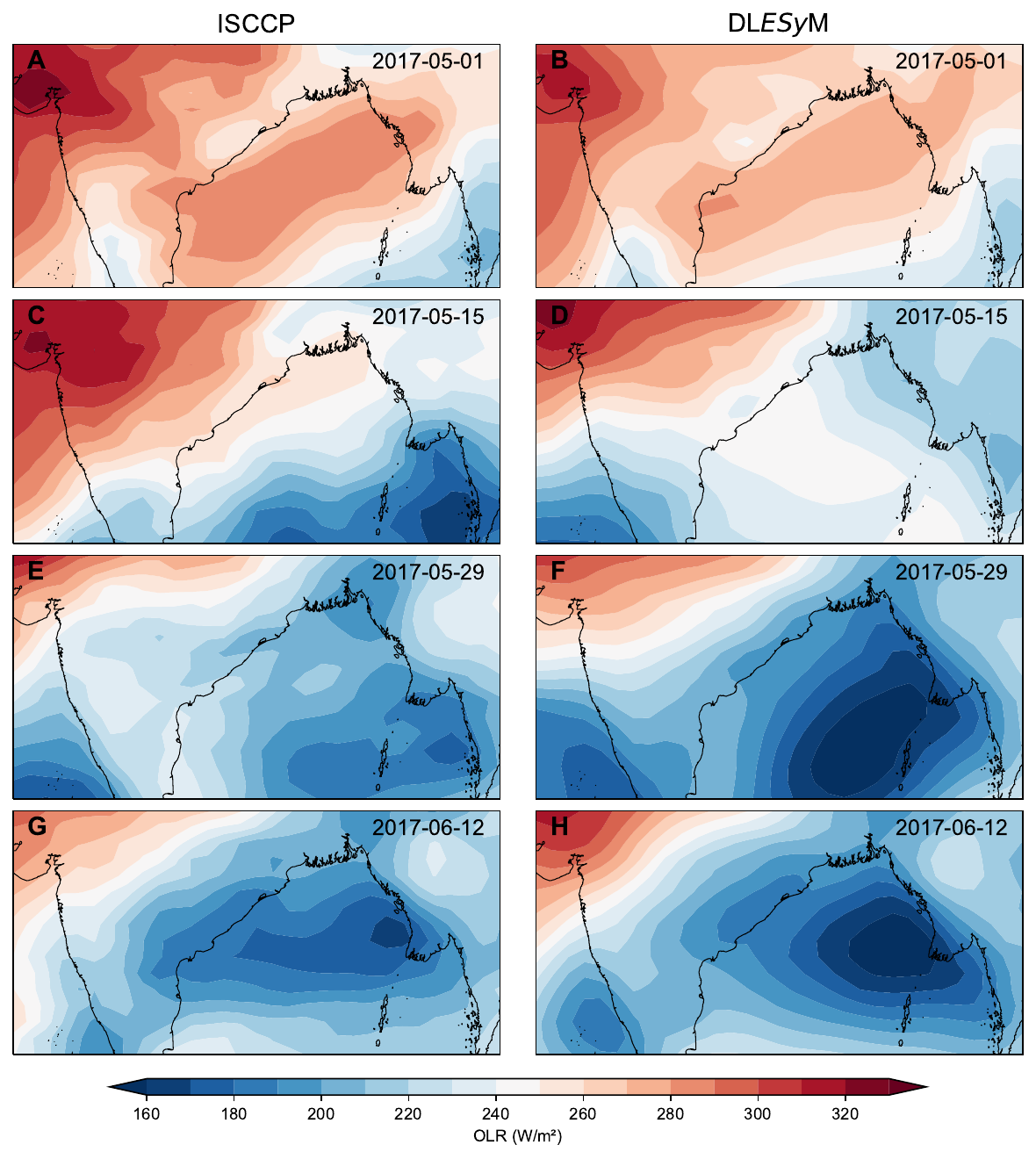}
  \caption{\textbf{Extended ISM OLR forecast with DL\textit{ESy}M.} May--June 2017 progression of 2-week-averaged OLR for the period beginning on the date noted in each panel in (A), (C), (E), (G): ISCCP data, and (B), (D), (F), (H): 6-month DL{\it ESy}M forecast from January 1, 2017.} 
  \label{fig:ISM_2D_2week}
\end{figure}

\begin{figure} 
  \centering  \includegraphics[width=0.9\textwidth]{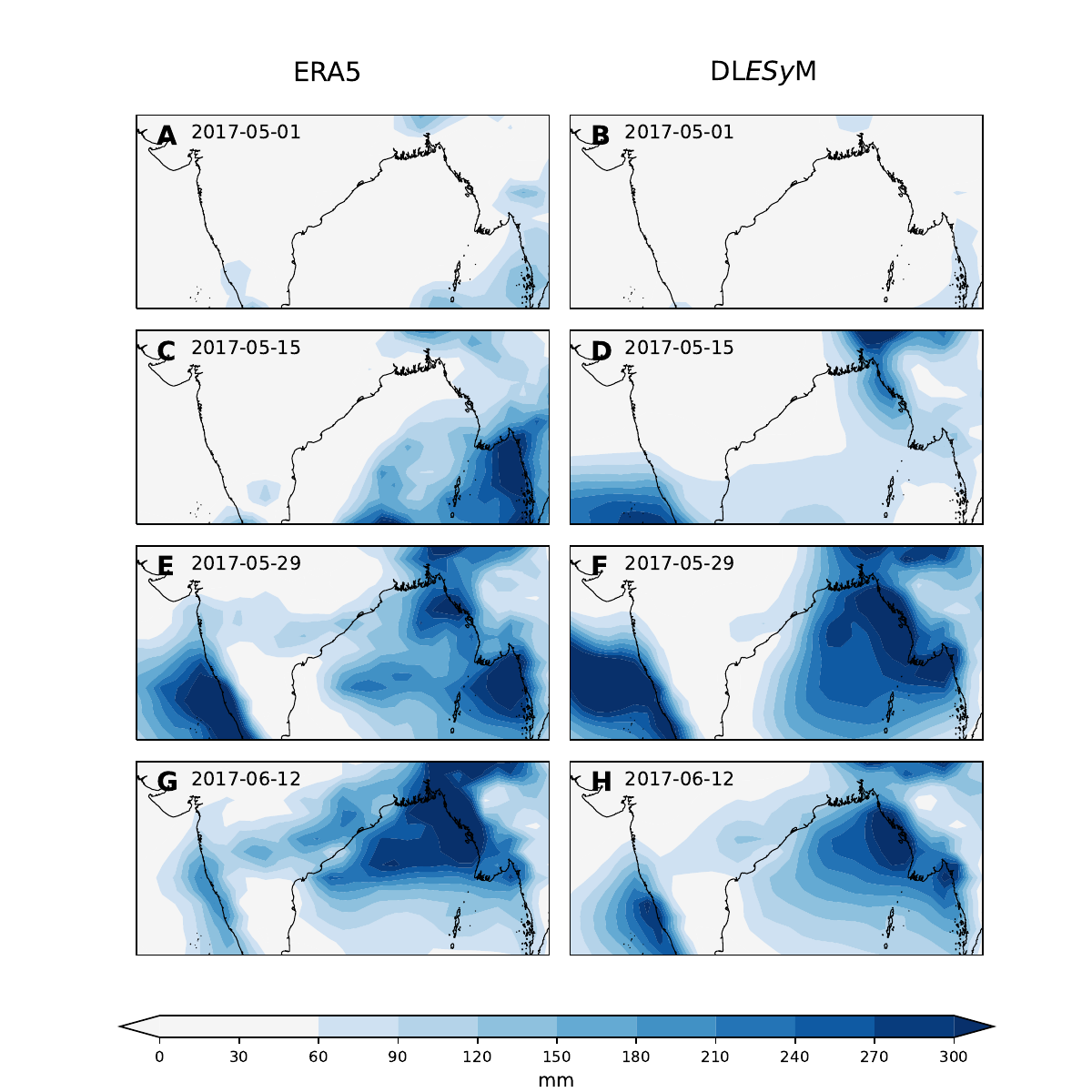}
  \caption{\textbf{Extended ISM precipitation forecast with DL\textit{ESy}M.} May--June 2017 progression of 2-week accumulated precipitation fields for the period beginning on the date noted in each panel in (A), (C), (E), (G): ERA5, and (B), (D), (F), (H): 6-month DL{\it ESy}M forecast from January 1, 2017. Considering the six-month forecast lead time, the predicted and analyzed fields are in reasonably good agreement in three of the four 2-week periods, with substantial differences only during May 15-28. }
  \label{fig:monsoon_precip}
\end{figure}

\begin{figure} 
  \centering  \includegraphics[width=0.7\textwidth]{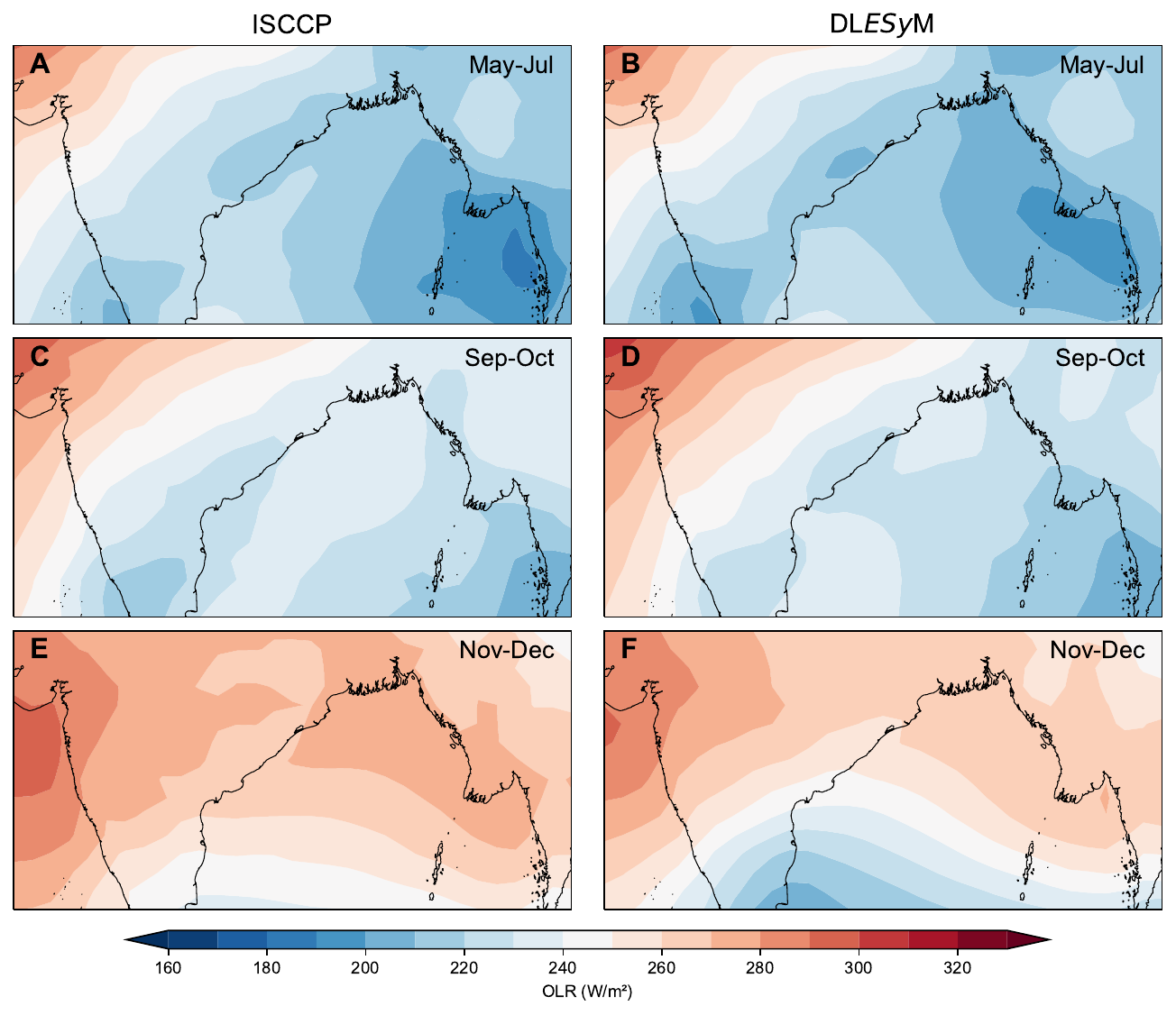}
  \caption{ \textbf{Climatological evolution of Indian Summer Monsoon (ISM) OLR in 100-year DL\textit{ESy}M simulation}. Annual cycle of multi-month averaged of OLR  in (A), (C), (E) ISCCP data for 1985-2014, and (B), (D), (F) years 2085-2114 in the DL{\it ESy}M rollout. Averaging period is indicated in each panel. During monsoon onset (May--July) and the beginning of it southward retreat (September--October), the DL{\it ESy}M fields are in close agreement with the observations.  In the November--December period, however, the OLR values are too low over southern India and the adjacent Indian Ocean, which leads to the errors at low-latitudes apparent in November and December in Fig.~\ref{fig:fig4}a,e.}
  \label{fig:monsoon_climo}
\end{figure}

Analyses reporting the Indian Summer Monsoon or ISM follow metrics defined in \cite{wangChoiceSouthAsian1999}. Outgoing longwave radiation values are spatially averaged between 10\textdegree\, and 25\textdegree\, North and 70\textdegree\, and 100\textdegree\, East to obtain the ISM index. Plotted time series in Fig.~\ref{fig:fig4}a are daily average values of this ISM Index. 

Here we examine DL{\it ESy}M's treatment of monsoon onset during May and June in a 6-month forecast beginning on January 1, 2017. 2-week averaged fields of OLR and 2-week accumulated precipitation are ploted in Figs. \ref{fig:ISM_2D_2week} and \ref{fig:monsoon_precip}, respectively. The OLR is observed by satellite, while the ERA5 precipitation in this region is generated by the ECMWF's global forecast model during each reanalysis cycle. Considering the six-month forecast lead time, the predicted and analyzed fields are in reasonably good agreement in three of the four 2-week periods, with substantial differences only during May 15-28.  

The climatological monsoon cycle has large amplitude. Indeed DL{\it ESy}M's ability to capture this climatological evolution (Fig. \ref{fig:monsoon_climo}) contributes to its apparent skill over the seasonal forecast. The extent to which DL{\it ESy}M can correctly capture interannual variations in monsoon onset in ensemble forecasts is a subject of current research.  Nevertheless, Figs.~\ref{fig:ISM_2D_2week} and \ref{fig:monsoon_precip} demonstrate that key atmospheric fields simulated by 
DL{\it ESy}M at the end of the 6-month rollout are physically realistic.

\end{document}